\documentclass[aps,pra,10pt,twocolumn,showpacs,preprintnumbers,amsmath,amssymb,nofootinbib,superscriptaddress,floatfix]{revtex4}

\usepackage{graphicx}
\usepackage{color}
\usepackage{float}
\usepackage[english]{babel}
\usepackage{blindtext}
\usepackage[T1]{fontenc}
\usepackage{balance}

\definecolor{dred}{rgb}{0.6,0.1,0}

\newcommand{\csixty}{\text{C}_{60}}
\newcommand{\wcmq}{\text{W}/\text{cm}^2}

\newcommand{\omlas}{\omega_\text{las}}
\newcommand{\llas}{I_\text{las}}

\newcommand{\bohr}{\mathrm{a}_0}

\newcommand{\be}{\begin{eqnarray}}
\newcommand{\ee}{\end{eqnarray}}

\begin{document}

\title{Forward-backward asymmetry of photoemission in $\csixty$
  excited by few-cycle laser pulses}
\author{C.-Z.~Gao}
\affiliation{Laboratoire de Physique Th\'eorique, Universit\'e de Toulouse, CNRS, UPS, France}
\author{P.~M. Dinh\footnote{corresponding author :  dinh@irsamc.ups-tlse.fr}}
\affiliation{Laboratoire de Physique Th\'eorique, Universit\'e de Toulouse, CNRS, UPS, France}
\author{P.-G.~Reinhard}
\affiliation{Institut f\"ur Theoretische Physik, Universit\"at Erlangen, Staudtstra\ss e 7, D-91058 Erlangen, Germany
}
\author{E.~Suraud}
\affiliation{Laboratoire de Physique Th\'eorique, Universit\'e de Toulouse, CNRS, UPS, France}
\author{C.~Meier}
\affiliation{Laboratoire Collisions-Agr\'egats-R\'eactivit\'e, Universit\'e de Toulouse, CNRS, UPS, France}

% \date{\today / Received: date / Revised version: date}

\date{\today}

\begin{abstract}
We theoretically analyze angle-resolved photo-electron spectra
(ARPES) generated by the interaction of $\csixty$ with
intense, short laser pulses.  In particular, we focus on the impact 
of the carrier-envelope phase (CEP) onto the angular distribution.
The electronic dynamics is described by time-dependent density
functional theory, and 
the ionic background of $\csixty$ is approximated by a
  particularly designed jellium model. Our results show
a clear dependence of the angular distributions onto the CEP
for very short pulses covering only very few laser cycles, which 
disappears for longer pulses. 
For the specific laser parameters used in a recent experiments, 
a very good agreement is obtained. 
Furthermore, the asymmetry is found 
to depend on the energy of the emitted photoelectrons. 
The strong influence of the angular asymmetry of electron emission 
onto the CEP and pulse duration 
suggests to use this sensitivity as a means to analyze the structure of few-cycle laser
pulses. 
\end{abstract}

\maketitle

\section{Introduction}
\label{sec:intro}

With the advance in laser technology, it has become possible to
generate femtosecond laser pulses which cover only few optical
cycles~\cite{Nis97,PauG01b}, and this has rapidly found a broad range
of applications in many disciplines, such as the generation of
attosecond pulses and precision control of chemical process. In this context,
one of the most intriguing aspects is that these extremely short pulses
may give access to time-resolved electronic dynamics in
atoms and molecules.
This paves the way to a multiplicity of interesting phenomena, such as
high-order harmonic generation (HOHG), above-threshold ionization (ATI),
and laser-induced molecular fragmentation, as has been seen already
in several earlier experimental and theoretical studies~\cite{Tem99,PauG01b,Bal03,Che05,Kli06,Liu11,Xie12,Sua15},
for reviews see Refs.~\cite{Bra00,Bec02,Mil06,Hae11}. It has also been
shown that structural information of the target can be retrieved using
short light pulses ~\cite{Mor08,Kan10}.

The carrier-envelope phase (CEP) is the phase of the fast oscillations of
the laser field relative to its envelope. For few-cycle lasers, this CEP
becomes a decisive laser parameter because the CEP offsets
modify the pattern of the pulse dramatically, which, in turn, can have
a strong impact on laser-induced electron dynamics. For example,
photoelectron emission induced by few-cycle laser fields can be
controlled by the CEP, leading to a pronounced forward-backward
(also called ``right-left'') asymmetry in the photoelectron spectra (PES).
This has been experimentally reported in~\cite{Pau03} where it was
found that the outcome depends on the photoelectrons' kinetic energy.
The energy dependence has been explained within a semiclassical model
by different electron emission processes in the low- and high-energy
regimes~\cite{Cor93}. In the low-energy regime, electrons are directly
emitted with a kinetic energy of up to the $2U_p$, where
$U_p=\llas/4\omlas^2$ (in atomic units) is the ponderomotive
energy of the laser field.

In the high-energy regime, electron recollision with the
  target system dominates, forming a plateau-like structure in
  the PES delimited by a well-defined cutoff.
The rescattered electrons can be accelerated to energies of roughly up
to the $10U_p$~\cite{Pau94} provided that the tail of the laser field
is still sufficiently high which is, however, rather critical for
few-cycle laser fields. Interestingly, a couple of theoretical
calculations have shown that the parts of PES related to electron
  recollisions are more sensitive to CEP than those parts
  related to electrons emitted
  directly~\cite{Mil03,Che05,Ton06,Qin08,Sua15}.
  Experimentally, the dependence of high-energy PES on the CEP has
been explored for atoms, e.g., xenon~\cite{Pau03,Kli08b}, argon~\cite{Lin05}, krypton~\cite{Kli08b}, as well as for small-sized dimer molecules, e.g., N$_2$
and O$_2$~\cite{Gaz11}. Recently, it has been studied in solids, such as
tungsten~\cite{Kru11} and gold nanotips~\cite{Par12} which were found to be
efficient and controllable nanoemitters of extreme ultraviolet (XUV) electrons,
thus allowing to investigate ultrafast electron dynamics in solids at an
attosecond time scale, for a recent review see Ref.~\cite{Kru12a}.

Compared to atoms and dimers, the $\csixty$ fullerene is a typical
example for a large system on the way from molecules to solids.
Thus the study of $\csixty$ might help toward understanding dynamical
properties of nanosystems. The advantage of $\csixty$ is its stability and
accessibility which renders it a useful laboratory for the study, e.g., of
thermal electron emission, charge migration, fragmentation channels,
HHG, and ATI, see Refs.~\cite{Her05,Cam06,Gan11,Lep15}. From a
geometrical point of view, $\csixty$ is similar to the outermost part 
of capped-carbon nanotips which are promising XUV electron nanoemitters.
In a previous
work~\cite{Gao16}, we have theoretically investigated the PES of $\csixty$
in strong fields using a near-infrared laser pulse ($\lambda_\mathrm{las}$=912 nm).
In that study, CEP effects were neglected 
since we considered comparatively long pulses comprising 
about 8 optical cycles. Recently,  
experiments on $\csixty$ using intense few-cycle
infrared laser pulses (720 nm and 4 fs)~\cite{Li15} were reported, 
in which a dramatic dependence of the PES on CEP was observed and qualitatively
reproduced by Monte Carlo (MC) and Quantum Dynamical (QD) simulations.

The aim of this article is to study the dependence of the PES on the CEP
for $\csixty$ illuminated by intense, infrared, few-cycle laser pulses
in a fully quantum-mechanical framework. Our modeling is based on Time-Dependent
Density-Functional Theory~\cite{Run84} with the time-dependent
local-density approximation using the jellium approximation for
the ionic background~\cite{Gao16}. We will focus on the dependence
of the CEP effect on pulse length and on the forward-backward asymmetry
of photoemission due to electron rescattering.

The paper is outlined as follows, Section~\ref{sec:model} briefly
describes the theoretical approach and the numerical analysis. Results
are presented and analyzed in Section~\ref{sec:re}. Finally, conclusions are
summarized in Section~\ref{sec:cons}.

\section{Formal framework}
\label{sec:model}

%%\subsection{Modeling}
%%\label{subsec:fram}

We describe the electronic dynamics of $\csixty$ by time-dependent
density functional theory (TDDFT) at the level of the time-dependent
local density approximation (TDLDA)~\cite{Dre90} using the
exchange-functional from~\cite{Per92}.  For an appropriate modeling of
electron emission, we augment TDLDA by a self-energy correction (SIC)
\cite{PeZ81}. As a full SIC treatment is computationally cumbersome
\cite{Mes08b}, we use it in a simplified, but reliable and efficient version
as an average density SIC (ADSIC)~\cite{Leg02}. The ADSIC suffices to put the
single-particle energies into right relation to continuum threshold such
that the ionization potential (IP) is correctly reproduced in a great variety
of systems~\cite{Klu13a} from simple atoms to large organic molecules.
In this context, a correct description of IP is crucial for photoemission
excited by external fields, in particular by strong fields as it
is assumed to be dominated by electrons in the highest occupied molecular
orbitals (HOMO)~\cite{Mul96}.

The ionic background (here carbon ions) is modeled within
the jellium approximation by a sphere of positive charge with a void at the center~\cite{Pus93,Bau01,Cor03}.  The jellium potential reads (in atomic units)~:
\begin{subequations}
\label{eq:psrho}
\begin{eqnarray}
  \upsilon_\mathrm{jel} (\mathbf r)
  &=&
  -\int \textrm d^{3} \mathbf r'
  \frac{ \rho_\mathrm{jel}(|\textbf{r}'|) } {|\textbf{r}-\textbf{r}'|}
  +
  \upsilon_\mathrm{ps}(|\mathbf r|)
  \;,
\label{eq:vjel}
\\
  \rho_\mathrm{jel} (r)
  &=&
  \rho_0 \,g(r)
  \;,
\label{eq:pspot}
\\
 \upsilon_\mathrm{ps} (r)
  &=&
  \upsilon_0 \, g(r)
  \;,
\label{eq:WS}
\\
   g(r)
   &=&
   \frac{1}{ \displaystyle 1+e^{(r\!-\! R_-)/\sigma}}\,
   \frac{1}{ \displaystyle 1+e^{(R_+\!-\!r)/\sigma}}
   \;,
\label{eq:gfunc}
\\
   R_\pm
    &=&
    R\pm\frac{\Delta R}{2}
    \;.
\label{eq:Rpm}
\end{eqnarray}
\end{subequations}
Here $g$ denotes the Woods-Saxon profile, providing a soft
  transition from bulk shell to the vacuum. The jellium potential
  $(\upsilon_\mathrm{jel}$) is augmented by an additional potential
  $\upsilon_\mathrm{ps}$ which is tuned to obtain reasonable values of
  the single-particle energies~\cite{Rei13}.  The shell radius $R$ is
  taken from experimental data as $R=6.7~\bohr$~\cite{Hed91}.  The
  other parameters are the same as those in Ref.~\cite{Gao16}.  With
  the present scheme, we reproduce rather well the electronic
  properties of $\csixty$: an IP at $E_\mathrm{IP}=7.62$~eV, a
  HOMO-LUMO gap of 1.77~eV, and a reasonable description of the
  photo-absorption spectrum~\cite{Rei13}. This is in nice agreement
  with experimental values~\cite{Lic91b,Sat10}.

It should be noted that the bulk density $\rho_0$ is
determined such that
$\int\textrm{d}^3\mathbf{r}\,\rho_\mathrm{jel}(\mathbf{r})~=N_\mathrm{el}=238$.
Note that this number of electrons is different from 240 for a real
$\csixty$. This is because no jellium model so far manages to place
the electronic shell at $N_\mathrm{el}=240$ as it should be.  Most
have the closure at $N_\mathrm{el}=250$ \cite{Pus93,Bau01,Cor03}.  The
present model with soft surfaces comes to $N_\mathrm{el}=238$ which is
much closer to the reality. Nevertheless, we have to keep in mind that
a jellium model is a rough approximation to a detailed ionic
structure. But it is a powerful approximation as it allows to
appropriately describe many features of electronic structure and
dynamics in solids~\cite{Ash76,Lem03} and cluster physics, see
Refs.~\cite{Kre93,Bra93}.  Recently, the present model has been
validated as one of efficient and reliable tools to describe electron
recollisions in strong fields in $\csixty$, see Ref.~\cite{Gao16}. The
jellium model stands naturally for a frozen ionic background. It is
justified for the present study 
where the laser pulses considered are so short that the nuclear 
dynamics can be neglected. 

\begin{figure}[htbp!]
 \centering
 \includegraphics[width=\linewidth]{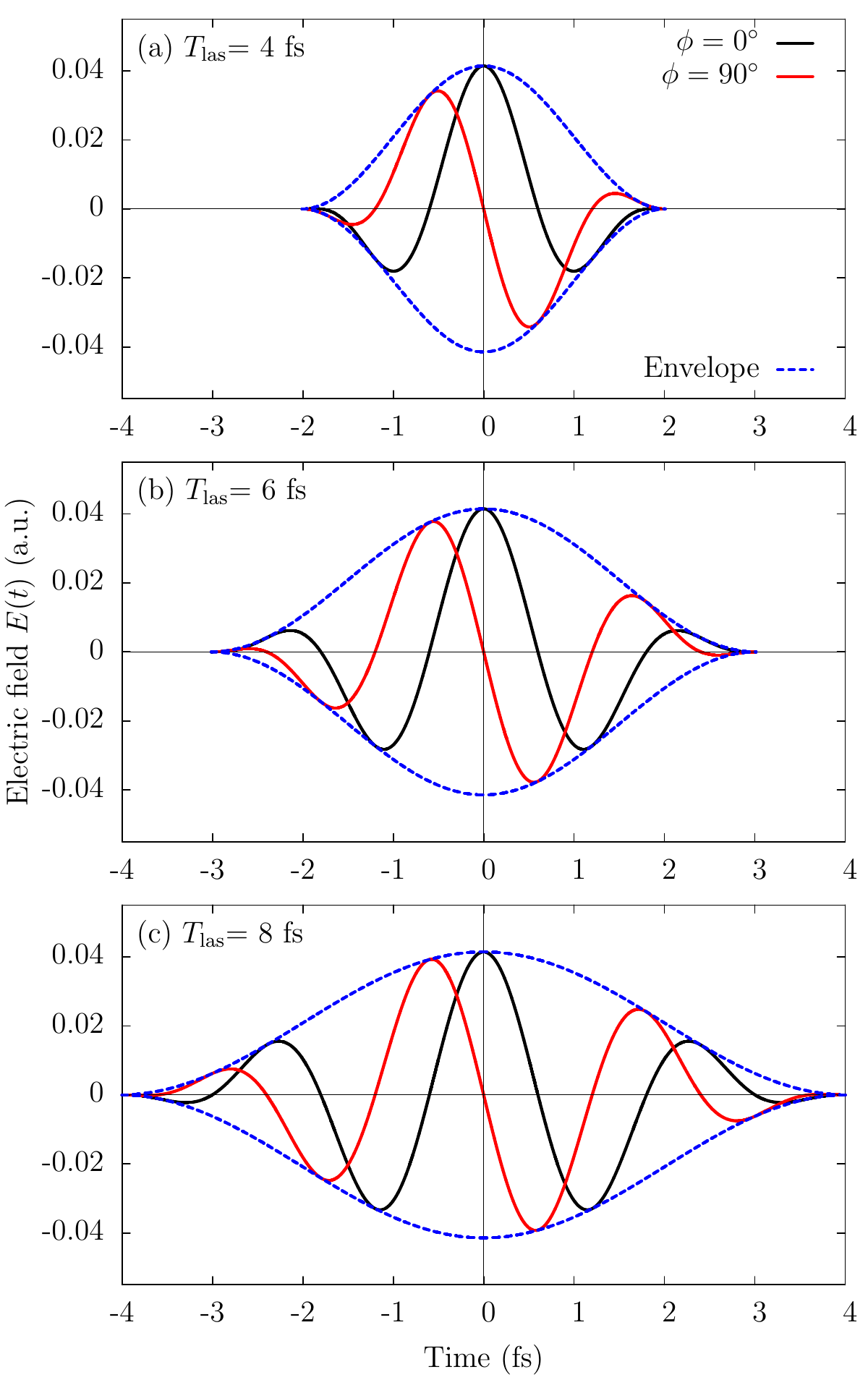}
 \caption{\label{fig:schematic} (Color online) The temporal part of
   the laser field (\ref{eq:laser}) for $T_\mathrm{las}$=4 fs (a), 6
   fs (b), 8 fs (c). Each panel is plotted at
   $\phi_\mathrm{CEP}=0^{\circ}$ (black lines) and $90^{\circ}$ (red
   lines). The envelope is shown in dashed blue lines. Horizontal ($E$=0) and
   vertical ($t$=0) solid lines are depicted to facilitate the CEP comparison.
   Other laser parameters are given in the text.}
\end{figure}

Within the dipole approximation, and assuming a linearly polarized 
laser pulse with the polarization vector along the $z$-axis $\mathbf{e}_z$,
the interaction with the 
laser field (in atomic units) is given by
\begin{eqnarray} 
v_{las}(r,t) & = &  E(t) \, \mathbf{r} \cdot \mathbf{e}_z
\end{eqnarray}
with the electric field chosen to be
\begin{eqnarray}
\label{eq:laser}
E(t) & = & E_0 \cos^2\left(\frac{\pi t}{T_\mathrm{las}}\right) \cos(\omlas t+\phi_\mathrm{CEP}) 
\end{eqnarray}
for $-T_\mathrm{las}/2 \leq t \leq T_\mathrm{las}/2$. 
Here, $E_0$ denotes the peak electric field, 
$\omlas$ the
carrier frequency, and $T_\mathrm{las}$ the total pulse
duration.
The CEP is comprised in the parameter
$\phi_\mathrm{CEP}$ which defines the phase between oscillations at
frequency $\omlas$ and the maximum of the $\cos^2$ envelope.
In what
follows, we use laser parameters close to those in recent
experiments~\cite{Li15}: laser frequency $\omlas=1.72$ eV (a
wavelength of 720 nm), intensity $I=6\times10^{13}~\wcmq$,
corresponding to field amplitude $E_0=1.1$ eV/a$_0$, and total
duration $T_\mathrm{las}$=4 fs, 6 fs, and 8 fs, corresponding to 1.7,
2.5, and 3.3 optical cycles (1 optical cycle = 2.4 fs). Note that
these laser parameters are associated with a ponderomotive energy
$U_p=2.9$ eV.
Figure~\ref{fig:schematic}
illustrates the temporal part of the laser field for the
three $T_\mathrm{las}$ under consideration, each one for the CEP
at $0^{\circ}$ (black curves) and $90^{\circ}$ (red curves).  
For $\phi_\mathrm{CEP}=0^{\circ}$, the center of the envelope
coincides with a maximum of the oscillations, while for
$\phi_\mathrm{CEP}=90^{\circ}$, it is shifted to match with the
nodal points of the electric field. Clearly, we see a substantial
change of the electric field due to the different CEP in each case. 

\subsection{Numerical details}
\label{subsec:numerics}

The TDLDA equations are solved numerically on a cylindrical grid
in coordinate space~\cite{Mon95a}. The static iterations towards
the electronic ground state are done with the damped gradient
method~\cite{Rei82} and time evolution employs the time-splitting
technique~\cite{Fei82}. For details of the numerical methods,
see~\cite{Cal00,Rei03a,Wop15}.
We use a numerical box which extends $500~\bohr$
in $z$ direction (along the laser polarization) and $250~\bohr$ orthogonal to it
(radial $r$ coordinate), with a grid spacing of $0.5~\bohr$ in both directions.
Time propagation is followed up to after 44 fs with a small time step
of $10^{-3}$ fs. Box size and time span are sufficiently large to
track completely the rescattering of electrons in the laser field
(ponderomotive motion). To account for ionization, absorbing
boundary conditions are implemented using a mask function~\cite{PGR06}.
The absorbing margin extends over 35 $\bohr$ (70 grid points) at each side.

The central observable of electron emission in our analysis are
angle-resolved photoelectron spectra (ARPES), i.e., the yield of emitted
electrons [$\mathcal{Y}(E_\mathrm{kin},\theta)$] as function of
kinetic energy $E_\mathrm{kin}$ and emission angle $\theta$. We
calculate ARPES by recording 
at each time step the single-electron wave functions \{$\psi_j(t,\mathbf{r}_{\mathcal{M}})$,
$j=1, \ldots, N_\mathrm{el}$\} at selected measuring points
$\mathbf{r}_{\mathcal{M}}$ near the absorbing layer and finally
transforming this information from time- to frequency-domain,
see~\cite{Poh01,DeG12,Din13,Dau16}. Finally, the PES is written as
\begin{equation}
  \mathcal{Y}(E_\mathrm{kin},\theta) 
  \propto 
  \sum_{j=1}^{N_{\mathrm{el}}}
 \lvert \widetilde{\psi_{j}}(E_\mathrm{kin},\mathbf r_{\mathcal M})
 \rvert^2 
\label{eq:pes}
\end{equation}
where $\widetilde{\psi_{j}}$ are the transformed wave functions in
energy domain. In case of strong fields, as we encounter here, 
the $\widetilde{\psi_{j}}$ are to be augmented by a phase factor
accounting for the ponderomotive motion, for technical details see 
\cite{Din13}.
The angle $\theta$ is defined with respect to $\mathbf{e}_z$, 
i.e. $\theta=0^{\circ}$ means electronic emission in the direction of $\mathbf{e}_z$. 
A detailed ARPES analysis requires a fine resolution. To that end, we
use an increment of 0.04 eV in energy and $1^{\circ}$ for
  the angular bins.

\section{Results and discussions}
\label{sec:re}

\subsection{CEP-averaged PES}

\begin{figure}[tbp!]
 \centering
 \includegraphics[width=\linewidth]{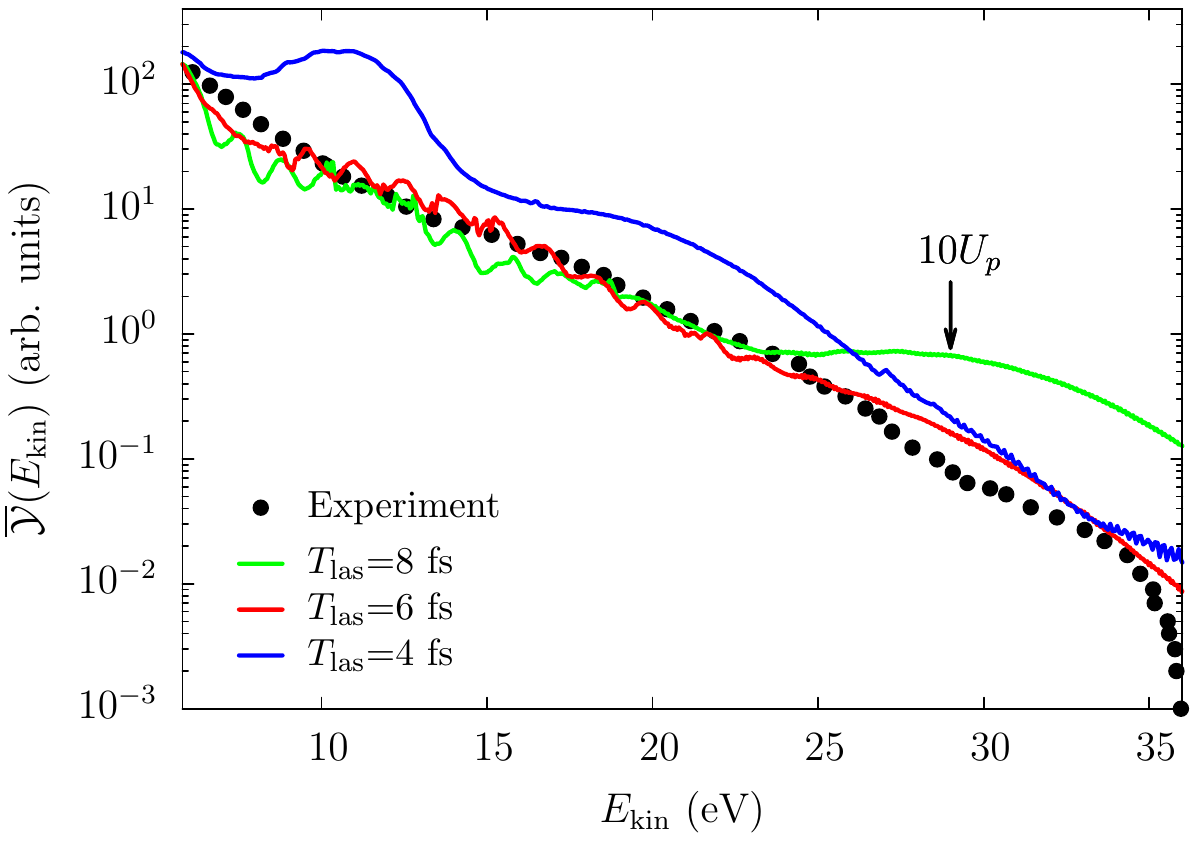}
 \caption{\label{fig:avrpescep} (Color online) CEP-averaged
   photoelectron spectra (PES) of $\csixty$ excited by intense,
   linearly polarized, few-cycle laser pulses. Experimental
   data~\cite{Li15} are shown as black circles and calculated
   results as full curves for different pulse lengths
   as indicated. The PES are collected in a cone with opening angle of
   $15^\circ$ at a fixed CEP and then averaged over a CEP range of
   $0^\circ$-$360^\circ$ at a step of $\Delta
   \phi_\mathrm{CEP}$=$15^\circ$. The high-energy cutoff
   position at $10 U_p$ is indicated by a black
   arrow.}
\end{figure}

We first look at CEP-averaged photoelectron spectra, simply denoted
by PES, of C$_{60}$ in a forward emission cone as measured
in the experiments of~\cite{Li15}. The computed PES are thus averaged
over CEP in a range of $0^\circ$-$360^\circ$ with $\Delta \phi_\mathrm{CEP}=15^\circ$
and collected in a forward cone with opening angle of $15^\circ$.
Figure~\ref{fig:avrpescep} shows the calculated results for the three pulse
lengths together with the experimental results (black solid circles)~\cite{Li15}.  
The patterns of the CEP-averaged PES are found to depend sensitively on
the pulse length $T_\mathrm{las}$.  At low
energies ($6\leq E_\mathrm{kin}\leq25$ eV), the photoelectron
yield decreases with increasing $T_\mathrm{las}$. It is interesting
to note that the PES for $T_\mathrm{las}$=4 fs reaches a maximum
around 10 eV. This coincides approximately with 3.2$\mathrm{U}_p$
which is the maximal energy upon the first return of rescattered
electrons. A similar pattern has been obtained in C$_{60}$ by quantum
dynamical (QD) calculations~\cite{Li15} under the same laser conditions.
For the longest pulse considered, $T_\mathrm{las}$=8 fs, we
find pronounced peaks which are the ATI peaks separated by the
photon energy. For shortest pulse lengths, these structures vanish
because the pulse does not have sufficient energy resolution any more.
In the high-energy regime ($25\leq E_\mathrm{kin}\leq36$ eV), we see a
reverse dependence on $T_\mathrm{las}$, where the longest pulse
($T_\mathrm{las}$=8 fs) leads to the highest yield
because there is more time to accelerate emitted electrons in the
still ongoing laser field. The most satisfactory agreement between TDLDA
results and experimental data is found
for $T_\mathrm{las}$=6 fs, which can nearly reproduce the
measured PES data in the full energy range. This strong
dependence of PES on pulse length may provide an opportunity to
characterize the experimental pulse duration by comparing the pattern
of PES to calculated results.

\subsection{Angle-resolved PES (ARPES)}

In a next step, we analyze the full ARPES at fixed CEP values for the
pulse length $T_\mathrm{las}$=6 fs where computed CEP-averaged PES
agree best with experiments, as seen in the previous section.
\begin{figure}[tbp!]
 \centering
 \includegraphics[width=\linewidth]{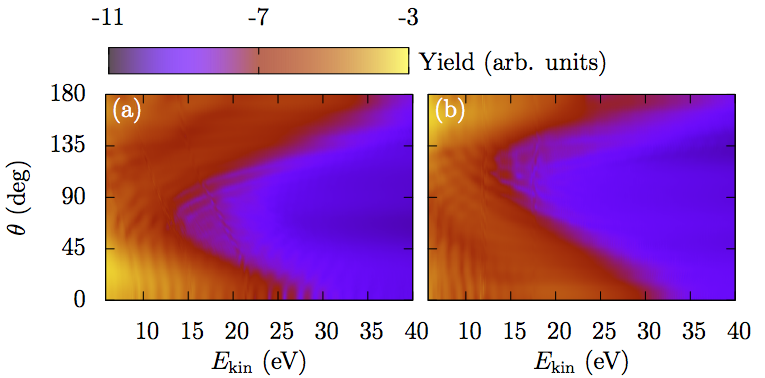}
 \caption{\label{fig:pespad6fs} (Color online) ARPES of C$_{60}$
   excited by a pulse with duration $T_\mathrm{las}$=6 fs, frequency
   $\omlas=1.72$ eV, and intensity $\llas=6\times10^{13}~\wcmq$ for 
   $\phi_\mathrm{CEP}=0^{\circ}$ (a) and $90^{\circ}$ (b).}
\end{figure}
Figure~\ref{fig:pespad6fs} shows ARPES for two typical CEP values,
$\phi_\mathrm{CEP}=0^{\circ}$ in panel (a) and $90^{\circ}$ in (b).
The most prominent feature in both cases is the remarkable 
$\theta=0^{\circ}\leftrightarrow 180^{\circ}$
asymmetry of the PES, particularly for electrons in the high-energy
regime ($E_\mathrm{kin}\geq25$ eV). At $\phi_\mathrm{CEP}=0^{\circ}$
in Fig.~\ref{fig:pespad6fs}(a), high-energy photoelectrons are
emitted favorably in the direction towards $\theta=180^\circ$,
which is characterized by a rather broad ramp extending to 35 eV.
In contrast, low-energy electrons ($6\leq E_\mathrm{kin}\leq15$ eV) are
emitted preferentially towards $\theta=0^\circ$. The observed
ARPES pattern is consistent with angular-integrated asymmetry maps
in experiments~\cite{Li15} for the CEP analyzed here. The preferences
of electron emission change for
  $\phi_\mathrm{CEP}=90^{\circ}$ in Fig.~\ref{fig:pespad6fs}(b), in which
  the low-energy electrons have more weight towards $\theta=180^\circ$
  while the high-energy electrons prefer the other way towards
  $\theta=0^\circ$. A similar asymmetry of the PES is also observed
for $T_\mathrm{las}=$4 fs and 8 fs (not shown). The example demonstrates that
strong few-cycle laser pulses can effectively control the ARPES.

This behavior has been observed also in experiments for
xenon atoms~\cite{Pau03} and has been theoretically analyzed
in~\cite{Mil03,Che05} based on the strong-field approximations 
and on the integration of three-dimensional time-dependent
Schr$\ddot{\rm o}$dinger equation. A semiclassical explanation
is that the generation of high-energy electrons originate from the
electron recolliding with the target, thus depending on the two time
instants at which electrons were released and scattered off,
respectively. It is difficult for few-cycle laser pulses to fulfill this condition
simultaneously in $\theta=0^{\circ}$ and $180^{\circ}$ directions.
However, it is possible to realize it in one of the
two directions by tuning the CEP offset of the laser field, as shown
in Fig.~\ref{fig:pespad6fs}. This has been first suggested
in~\cite{PauG01b} as a phase-meter to determine the absolute phase of
an ultrashort laser pulse. We shall show in the next step that such a
phase-meter strongly depends on pulse duration and that it becomes
invalid with increasing pulse length.

\subsection{Asymmetry versus CEP}

The dominant feature of the ARPES in Fig.~\ref{fig:pespad6fs} is the
strong influence of CEP on the energy resolved forward-backward
asymmetry. This was also found in the previous work where the asymmetry
often produces regular oscillations for the CEP as function of kinetic
energy~\cite{Mil06}. To investigate such oscillations for the
present example, we define the asymmetry $\eta$ as
\begin{equation}
  \eta(\phi_\mathrm{CEP})
  =
  \frac{\mathcal{Y}(E_{1-2},\theta^+)
           -\mathcal{Y}(E_{1-2},\theta^-)}
       {\mathcal{Y}(E_{1-2},\theta^+)
           +\mathcal{Y}(E_{1-2},\theta^-)}
\label{eq:eta}
\end{equation}
where $E_{1-2}$ denotes in brief the integration in the energy
interval $[E_1:E_2]$ and $\theta^+$ and $\theta^-$ stand for cones of
emission angles.  In most experiments, photoelectron yields are
collected in a cone angle of $15^{\circ}$. This means
that $\theta^+$ corresponds to $\vert\theta^+\vert\leq15^\circ$ and
$\theta^-$ to $\vert180-\theta^-\vert\leq15^\circ$. We use the same
convention for our theoretical analysis.

\begin{figure}[tbp!]
 \centering
 \includegraphics[width=\linewidth]{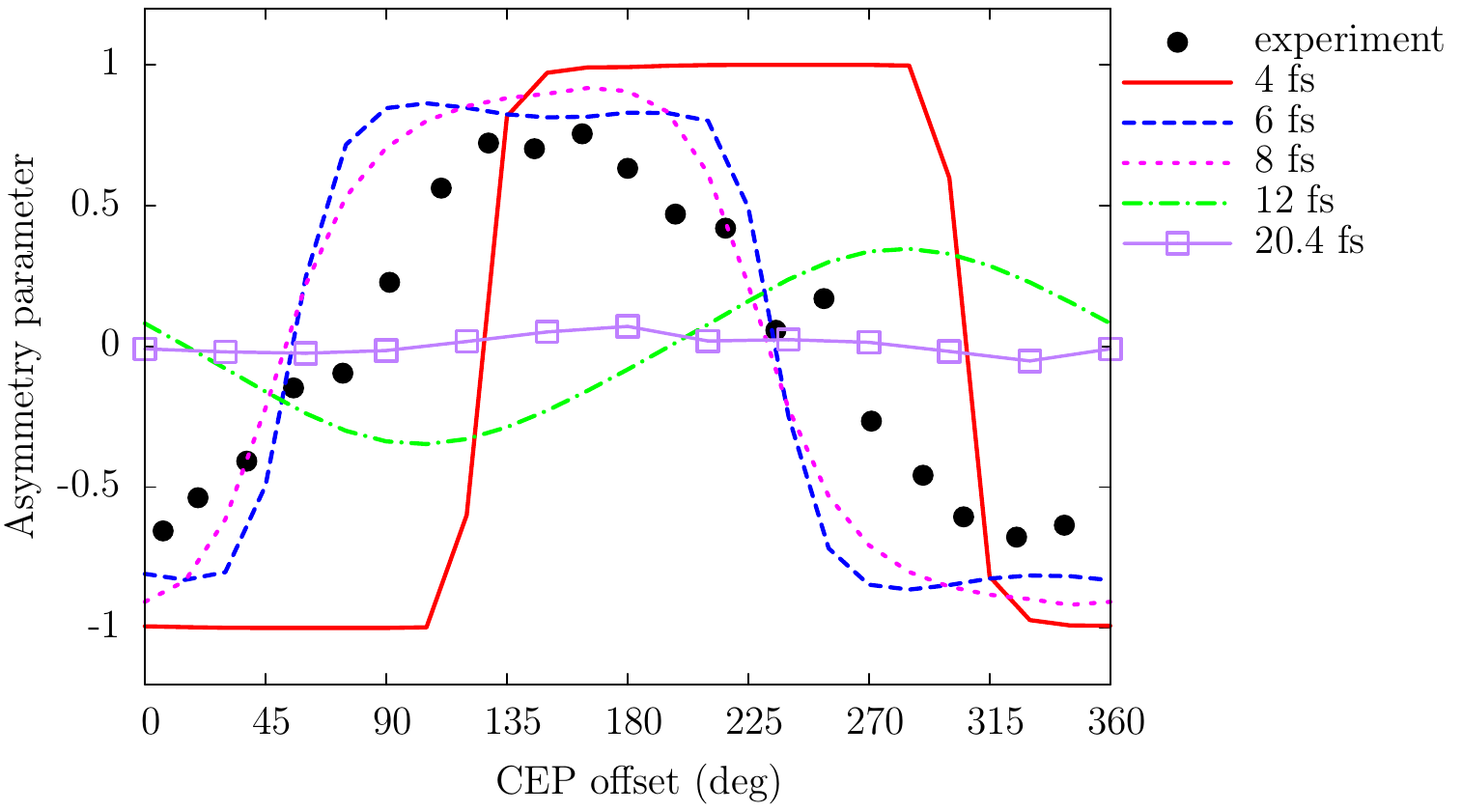}
 \caption{\label{fig:asys} (Color online) Comparison of the asymmetry
   parameter $\eta$, see Eq.~(\ref{eq:eta}), from experiments~\cite{Li15}
   (black solid circles) with theoretical results (red solid curves
   for $T_\mathrm{las}=$4 fs, blue dashes for 6 fs, magenta
   dots for 8 fs, green dash-dotted curves for 12 fs, and
   purple open squares for 20.4 fs). All cases
   have been integrated in the same range of kinetic energies
   $[E_1:E_2]=[23~\mathrm{eV}:30.6~\mathrm{eV}]$ as done in experiments.
 }
\end{figure}

Figure~\ref{fig:asys} compares the asymmetry parameter $\eta$ between
experiments (black solid circles)~\cite{Li15} and calculated results
for various pulse lengths $T_\mathrm{las}$. 
We find a good agreement
between experimental data and present results for $T_\mathrm{las}=$6
fs (blue dashes) as well as 8 fs (magenta dots) regarding the width and
position of the maxima of $\eta(\phi_\mathrm{CEP})$. Yet, slight
differences remain for the shape of$\eta(\phi_\mathrm{CEP})$.
The experimental curve
shows softer transitions than the two theoretical curves.  
The cos$^2$ envelope of the
theoretical pulse, see Eq.~(\ref{eq:laser}), is surely more confined than the
experimental pulse which is often assumed to have Gaussian
envelope, but may easily be plagued by prepulses~\cite{Giu06}.

The pattern of $\eta(\phi_\mathrm{CEP})$ differ substantially from the
experiment for other $T_\mathrm{las}$ (smaller than 6-8 fs and larger
ones) concerning position of maxima/minima, amplitude of oscillations,
and softness on $T_\mathrm{las}$. Clear trends are seen for the
amplitude which decreases with increasing $T_\mathrm{las}$ and the
softness which increases with $T_\mathrm{las}$, both together
eventually wiping out the signal for long pulses. The trends are
plausible. Very short pulse shrink basically to one oscillation and so
become extremely sensitive %\CZGfoot{I think it should be
%  insensitive. One oscillation means the emission is dominated either
%  in the forward or in the backward.\\ \PGR{A very short pulse
%    produces basically one single peak. This sign of this peak stays
%    constant for a range of CEP and changes suddenly at a certain
%    value of CEP. You can see this on the step like pattern of
%    asymmetry versus CEP for small $T_\mathrm{pulse}$. These rapid
%    changes is what I call sensitive. Longer pulses render the
%    transition smoother and reduce the amplitude. That is what I call
%    less sensitive.}}
to the CEP while more and more comparably high
oscillations in longer pulses render the CEP less crucial.
%\PGRfoot{Ionic motion may come into play for pulses longer than
%  10 fs. But this will only further soften the pattern. Should we
%  mention that? \\ \PGRmod{I propose not to mention ionic motion here.} \MD{I agree}}.  
This strong sensitivity of the signal
$\eta(\phi_\mathrm{CEP})$ to pulses parameters raises the
question how sensitive the result is to details of the pulse
profile. To check that, we have also run calculations with a Gaussian
envelope for the laser pulse instead of the cos$^2$ envelope used
above, see Eq. (\ref{eq:laser}). The results are practically the
same if the same FWHM is used. Therefore, we conclude that pulse
length is the decisive parameter and measuring $\eta(\phi_\mathrm{CEP})$
can give access to this parameter.  However, $\eta$ emerges from combined action of
laser pulses and responding system. This aspect, i.e., the influence of the
  system, will be addressed in future work.

Although it is plausible that the impact of CEP fades away for
longer pulses, CEP-dependent asymmetry of the PES can be recovered
also for longer pulses by a collinear, two-color pump-probe
scheme, namely, a combination of a fundamental laser ($\omega$) and
its $n$-th order harmonic ($n\omega$) typically represented by a $\omega$-$2\omega$
laser setup.
The presence of a second harmonic is used to twist the
  field strength of the fundamental mode by varying the delay
  phase, resulting in the asymmetry in the field amplitude
  influencing ionization as well as the rescattering. The
$\omega$-$n\omega$ scheme has previously been proposed
in~\cite{Pau95}, and has later been used to explore the PES asymmetry
in sodium clusters (Na$_4$ and Na$_4^+$) excited by intense 7-cycle
laser fields~\cite{Ngu04}. A more recent experimental application of $\omega$-$2\omega$
combined laser pulses on rare gas atoms and CO$_2$ molecule is found in
  Ref.~\cite{Skr15}. On this basis, the comparison of
  controlling efficiency of CEP-dependent asymmetry between
  one-color few-cycle fields and two-color multiple-cycle fields is
  an intriguing topic, yet this is beyond the scope of present study,
  thus we postpone it to the next exploration.
 
\begin{figure}[bp!]          
\centerline{\includegraphics[width=\linewidth]{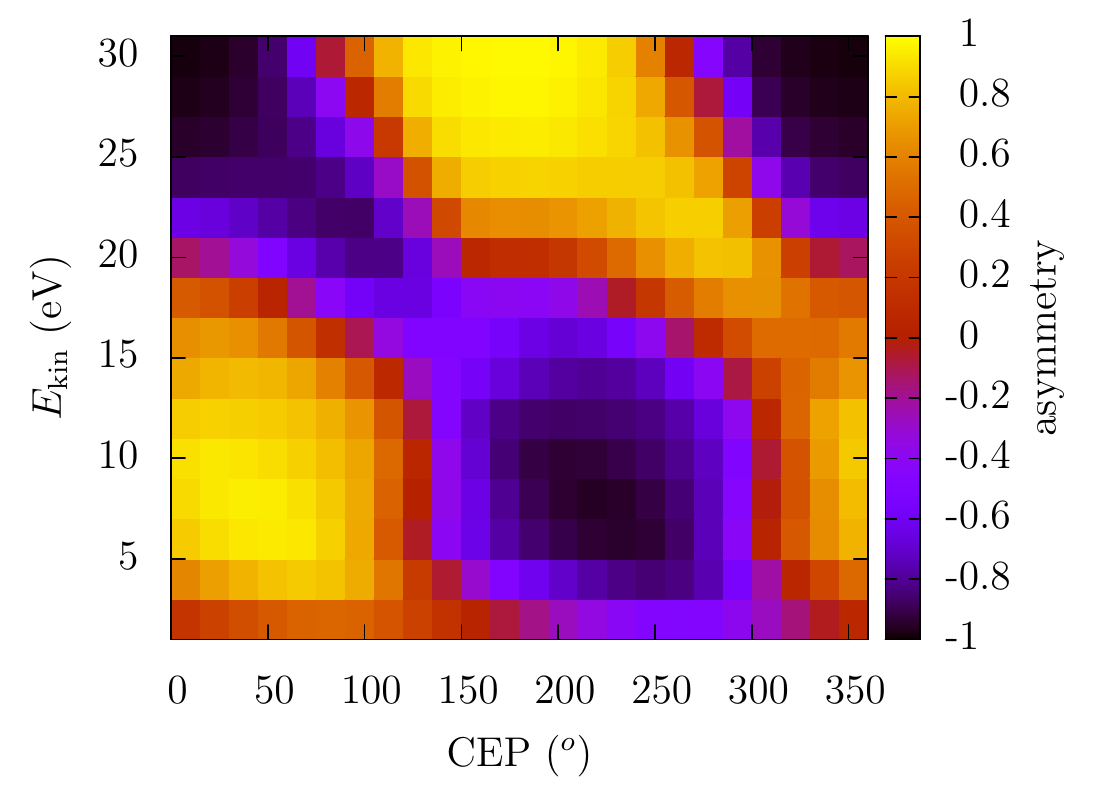}}
\caption{\label{fig:etaekin} (Color online) Dependence of the
  asymmetry parameter on the CEP and on the kinetic energy of 
the emitted electrons (for $T_\mathrm{las}=$6 fs).  A strong CEP 
dependence is found, with opposite behavior for 
low ($ < 15~\mathrm{eV}$ and high energy electrons (between $23~\mathrm{eV}$ and $31~\mathrm{eV}$).}
\end{figure}

Since the asymmetry parameter (\ref{eq:eta}) depends on the
kinetic energy of the photoelectrons, it is also interesting to study its
dependence on the energy. Figure~\ref{fig:etaekin} shows $\eta$ as
a function of the kinetic energy and the $\phi_\mathrm{CEP}$, using
$T_\mathrm{las}=$6 fs. 
As most striking result, we find the the anisotropy parameter strongly depends both 
on the CEP and the kinetic energy of the emitted photoelectrons. For high energy electrons, 
i.e, between $\sim 23$ and $\sim 30~\mathrm{eV}$, we find the strong CEP dependence already visible in Fig.~\ref{fig:asys}. 
For low energy electrons, those less than $\sim 15~\mathrm{eV}$, we find an opposite behavior. % comparing to the CEP-dependence of high-energy electrons.
This reflects the complex electron dynamics taking place during the interaction 
with the ultrashort laser pulse, which changes drastically for different CEP's. 

\begin{figure}[htbp!]
\centering
\includegraphics[width=\linewidth]{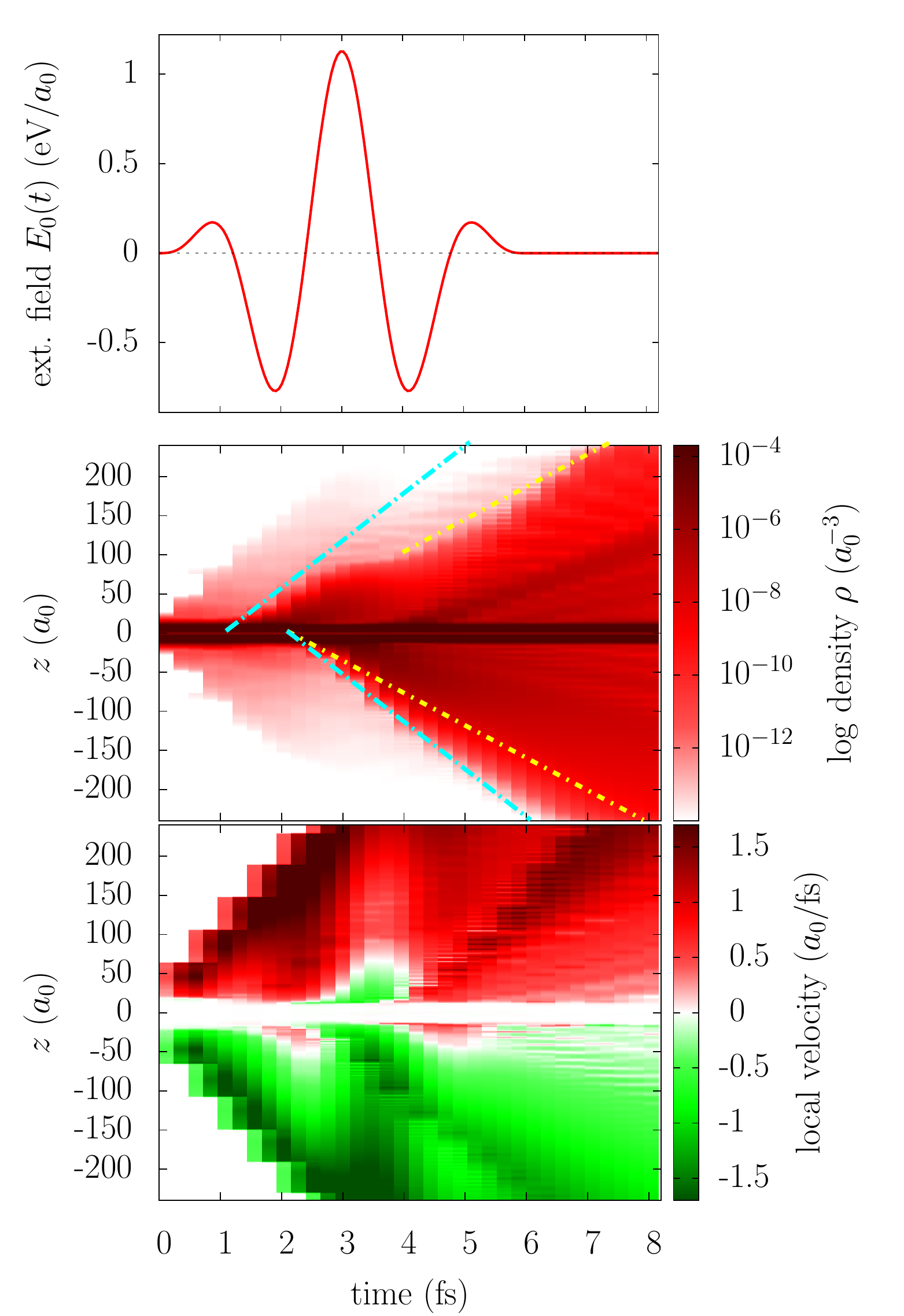}
\caption{\label{fig:tevol-CEP00} Time evolution of external field
  (upper panel), local density along $z$-axis (middle panel), and
  local velocity along $z$-axis (lower panel) for C$_{60}$ excited by
  a laser pulse with a duration of $T_\mathrm{las}$=6 fs, frequency
  $\omega_\mathrm{las}=1.72$ eV, intensity
  $I_\mathrm{las}=6\times10^{13}$W/cm$^2$, and CEP
  $\phi_\mathrm{CEP}=0^{\circ}$. 
%\CZG{The external field is normalized to
 % the maximum. For a better visualization, the density in the middle panel
%  is presented in logarithmic scale.} 
The cyan lines in the density plot
  (middle panel) indicate a velocity $\pm v=58$ a$_0$/fs corresponding
  to a kinetic energy of 13.6 eV and the yellow lines of $\pm v=41$
  a$_0$/fs corresponding to 27.2 eV.}
\end{figure}
To explain these features in more detail, we have analyzed the electron dynamics
in the case of $\phi_\mathrm{CEP}=0^\circ$. These results are depicted in Fig.~\ref{fig:tevol-CEP00}
It shows the time evolution of the density (middle panel) and
the velocity distribution (bottom panel), 
together with the electric field of the laser pulse (top panel). 
In analyzing the results, one has to concentrate on the
low-density tail corresponding to the finally emitted flow. The
velocities need to be looked together with the density because
velocity alone does not indicate the importance of a contribution
(which is weighted, of course, with the density).
As soon as the electric field sets on, both the density and the velocity
show oscillations synchronous with the external field. A bunch of high
positive velocities (cyan line with positive slope) develops followed by a
density shift to positive $z$ short after the first negative peak in
$E_0$ at 2 fs. This turns into opposite direction following the strong
positive peak in $E_00$ after 3 fs and a last swap back with the
counter peak after 4 fs.  A particularly interesting process takes
places with the largest peak at 3 fs. The negative peak before has
triggered a strong flow to positive $z$. This is abruptly stopped and
counter-weighted by the subsequent large positive field (exerting a
force in the negative direction). The positive flow recovers only with the
second negative $E_0$ peak and leaves the system with a kinetic energy
of about 13.6 eV (yellow line with positive slope). The large
positive peak, on the other hand, releases a bunch of fast electrons
towards negative direction which escapes eventually with the higher
kinetic energy of 27.2 eV (cyan line with negative slope). The detailed
time-resolved picture so nicely elucidates the peaks on $0^\circ$ and
$180^\circ$ direction in the previous figure.

\section{Conclusions}
\label{sec:cons}

We have investigated the impact of the carrier-envelope phase (CEP) of
very short laser pulses on the angle-resolved photo-electron spectra
(ARPES) of the $\csixty$ cluster.
To this end, we used as a tool
time-dependent density-functional theory at the level of the
time-dependent local-density approximation. This was augmented by
self-interaction correction to achieve a correct ionization potential
which is crucial for an appropriate description of photoemission dynamics.
The ionic background is assumed to be frozen during the femtosecond
dynamics. For  $\csixty$, it is approximated by a jellium model which is particularly
tuned to its special geometry. Absorbing boundary conditions
are used to describe electron emission and ARPES are computed with
sampling the time evolution of the single-electron wave functions at
selected measuring points close to the absorbing margins. 

Our results depend sensitively on the laser pulse length. This holds
already for the global signal of CEP-averaged photo-electron spectra
where we find a good agreement with experimental results when
using the appropriate pulse length (and huge deviations from data for
other pulse lengths). Particular attention was paid to the angular
asymmetry of ARPES. A short glance at the full ARPES distribution and
a detailed evaluation of asymmetry as a function of energy show that the
asymmetry behaves different in low- and high-energy regime. This is explained
in detail by analyzing the time-dependent density and velocity distributions
of the accelerated electrons. 

Following experimental data, we have focused then on the dependence of
asymmetry on CEP in the regime of high-energy emission.  For very
short pulses, we find a strongly varying function oscillating with
steep slopes between the forward/backward extremes of asymmetry.
These patterns change significantly with the pulse parameters. The
amplitude of oscillations shrinks with increasing pulse length while
the pattern become softer. In particular, the signal practically disappears for
longer pulses covering 8 laser cycles or more. We find again a good
agreement with experimental data for the appropriate pulse length, the
same which also allowed to reproduce the CEP-averaged photo-electron
spectra.

We have studied the sensitivity of the results to the detailed
profile of the laser pulse (Gaussian versus cos$^2$). The
differences are so small that pulse profiles cannot be identified
clearly from the asymmetry signal. 

This study emphasizes the amount of detailed information that can be gained 
from a systematic scan of the ARPES as a function of the CEP. For example,
the high sensitivity of asymmetry versus CEP to the laser pulse may be
used for an independent measurement of pulse parameters. 
This may constitute an interesting aspect for ultrashort 
pulse characterization. However, in order to 
develop the shown methodology in this direction, 
one needs to carefully disentangle pulse properties
from system properties, as resonances. 
Research along these lines are currently being undertaken. 

\section*{Acknowledgments:}
 
We thank Institut Universitaire de France, European ITN network CORINF
and French ANR contract LASCAR (ANR-13-BS04-0007) for support during
the realization of this work. One of authors (C.-Z.G.) is grateful for
the financial support from China Scholarship Council (CSC) (No.
[2013]3009). It was also granted access to the HPC resources of CalMiP
(Calcul en Midi-Pyr\'en\'ees) under the allocation P1238, and of RRZE
(Regionales Rechenzentrum Erlangen).\\

\bigskip
%set a link to that site
\bibliographystyle{apsrev}
\bibliography{sfeCEP.bib}

\begin{thebibliography}{67}
\expandafter\ifx\csname natexlab\endcsname\relax\def\natexlab#1{#1}\fi
\expandafter\ifx\csname bibnamefont\endcsname\relax
  \def\bibnamefont#1{#1}\fi
\expandafter\ifx\csname bibfnamefont\endcsname\relax
  \def\bibfnamefont#1{#1}\fi
\expandafter\ifx\csname citenamefont\endcsname\relax
  \def\citenamefont#1{#1}\fi
\expandafter\ifx\csname url\endcsname\relax
  \def\url#1{\texttt{#1}}\fi
\expandafter\ifx\csname urlprefix\endcsname\relax\def\urlprefix{URL }\fi
\providecommand{\bibinfo}[2]{#2}
\providecommand{\eprint}[2][]{\url{#2}}

\bibitem[{\citenamefont{Nisoli et~al.}(1997)\citenamefont{Nisoli, De~Silvestri,
  Svelto, Szip{\"o}cs, Ferencz, Spielmann, Sartania, and Krausz}}]{Nis97}
\bibinfo{author}{\bibfnamefont{M.}~\bibnamefont{Nisoli}},
  \bibinfo{author}{\bibfnamefont{S.}~\bibnamefont{De~Silvestri}},
  \bibinfo{author}{\bibfnamefont{O.}~\bibnamefont{Svelto}},
  \bibinfo{author}{\bibfnamefont{R.}~\bibnamefont{Szip{\"o}cs}},
  \bibinfo{author}{\bibfnamefont{K.}~\bibnamefont{Ferencz}},
  \bibinfo{author}{\bibfnamefont{C.}~\bibnamefont{Spielmann}},
  \bibinfo{author}{\bibfnamefont{S.}~\bibnamefont{Sartania}}, \bibnamefont{and}
  \bibinfo{author}{\bibfnamefont{F.}~\bibnamefont{Krausz}},
  \bibinfo{journal}{Opt. Lett.} \textbf{\bibinfo{volume}{22}},
  \bibinfo{pages}{522} (\bibinfo{year}{1997}).

\bibitem[{\citenamefont{Paulus et~al.}(2001)\citenamefont{Paulus, Grasbon,
  Walther, Villoresi, Nisoli, Stagira, Priori, and De~Silvestri}}]{PauG01b}
\bibinfo{author}{\bibfnamefont{G.}~\bibnamefont{Paulus}},
  \bibinfo{author}{\bibfnamefont{F.}~\bibnamefont{Grasbon}},
  \bibinfo{author}{\bibfnamefont{H.}~\bibnamefont{Walther}},
  \bibinfo{author}{\bibfnamefont{P.}~\bibnamefont{Villoresi}},
  \bibinfo{author}{\bibfnamefont{M.}~\bibnamefont{Nisoli}},
  \bibinfo{author}{\bibfnamefont{S.}~\bibnamefont{Stagira}},
  \bibinfo{author}{\bibfnamefont{E.}~\bibnamefont{Priori}}, \bibnamefont{and}
  \bibinfo{author}{\bibfnamefont{S.}~\bibnamefont{De~Silvestri}},
  \bibinfo{journal}{Nature} \textbf{\bibinfo{volume}{414}},
  \bibinfo{pages}{182} (\bibinfo{year}{2001}).

\bibitem[{\citenamefont{Tempea et~al.}(1999)\citenamefont{Tempea, Geissler, and
  Brabec}}]{Tem99}
\bibinfo{author}{\bibfnamefont{G.}~\bibnamefont{Tempea}},
  \bibinfo{author}{\bibfnamefont{M.}~\bibnamefont{Geissler}}, \bibnamefont{and}
  \bibinfo{author}{\bibfnamefont{T.}~\bibnamefont{Brabec}},
  \bibinfo{journal}{JOSA B} \textbf{\bibinfo{volume}{16}}, \bibinfo{pages}{669}
  (\bibinfo{year}{1999}).

\bibitem[{\citenamefont{Baltu{\v{s}}ka
  et~al.}(2003)\citenamefont{Baltu{\v{s}}ka, Udem, Uiberacker, Hentschel,
  Goulielmakis, Gohle, Holzwarth, Yakovlev, Scrinzi, H{\"a}nsch
  et~al.}}]{Bal03}
\bibinfo{author}{\bibfnamefont{A.}~\bibnamefont{Baltu{\v{s}}ka}},
  \bibinfo{author}{\bibfnamefont{T.}~\bibnamefont{Udem}},
  \bibinfo{author}{\bibfnamefont{M.}~\bibnamefont{Uiberacker}},
  \bibinfo{author}{\bibfnamefont{M.}~\bibnamefont{Hentschel}},
  \bibinfo{author}{\bibfnamefont{E.}~\bibnamefont{Goulielmakis}},
  \bibinfo{author}{\bibfnamefont{C.}~\bibnamefont{Gohle}},
  \bibinfo{author}{\bibfnamefont{R.}~\bibnamefont{Holzwarth}},
  \bibinfo{author}{\bibfnamefont{V.}~\bibnamefont{Yakovlev}},
  \bibinfo{author}{\bibfnamefont{A.}~\bibnamefont{Scrinzi}},
  \bibinfo{author}{\bibfnamefont{T.}~\bibnamefont{H{\"a}nsch}},
  \bibnamefont{et~al.}, \bibinfo{journal}{Nature}
  \textbf{\bibinfo{volume}{421}}, \bibinfo{pages}{611} (\bibinfo{year}{2003}).

\bibitem[{\citenamefont{Chelkowski and Bandrauk}(2005)}]{Che05}
\bibinfo{author}{\bibfnamefont{S.}~\bibnamefont{Chelkowski}} \bibnamefont{and}
  \bibinfo{author}{\bibfnamefont{A.~D.} \bibnamefont{Bandrauk}},
  \bibinfo{journal}{Phys. Rev. A} \textbf{\bibinfo{volume}{71}},
  \bibinfo{pages}{053815} (\bibinfo{year}{2005}).

\bibitem[{\citenamefont{Kling et~al.}(2006)\citenamefont{Kling, Siedschlag,
  Verhoef, Khan, Schultze, Uphues, Ni, Uiberacker, Drescher, Krausz
  et~al.}}]{Kli06}
\bibinfo{author}{\bibfnamefont{M.}~\bibnamefont{Kling}},
  \bibinfo{author}{\bibfnamefont{C.}~\bibnamefont{Siedschlag}},
  \bibinfo{author}{\bibfnamefont{A.~J.} \bibnamefont{Verhoef}},
  \bibinfo{author}{\bibfnamefont{J.}~\bibnamefont{Khan}},
  \bibinfo{author}{\bibfnamefont{M.}~\bibnamefont{Schultze}},
  \bibinfo{author}{\bibfnamefont{T.}~\bibnamefont{Uphues}},
  \bibinfo{author}{\bibfnamefont{Y.}~\bibnamefont{Ni}},
  \bibinfo{author}{\bibfnamefont{M.}~\bibnamefont{Uiberacker}},
  \bibinfo{author}{\bibfnamefont{M.}~\bibnamefont{Drescher}},
  \bibinfo{author}{\bibfnamefont{F.}~\bibnamefont{Krausz}},
  \bibnamefont{et~al.}, \bibinfo{journal}{Science}
  \textbf{\bibinfo{volume}{312}}, \bibinfo{pages}{246} (\bibinfo{year}{2006}).

\bibitem[{\citenamefont{Liu et~al.}(2011)\citenamefont{Liu, Liu, Deng, Wu,
  Jiang, and Gong}}]{Liu11}
\bibinfo{author}{\bibfnamefont{Y.}~\bibnamefont{Liu}},
  \bibinfo{author}{\bibfnamefont{X.}~\bibnamefont{Liu}},
  \bibinfo{author}{\bibfnamefont{Y.}~\bibnamefont{Deng}},
  \bibinfo{author}{\bibfnamefont{C.}~\bibnamefont{Wu}},
  \bibinfo{author}{\bibfnamefont{H.}~\bibnamefont{Jiang}}, \bibnamefont{and}
  \bibinfo{author}{\bibfnamefont{Q.}~\bibnamefont{Gong}},
  \bibinfo{journal}{Phys. Rev. Lett.} \textbf{\bibinfo{volume}{106}},
  \bibinfo{pages}{073004} (\bibinfo{year}{2011}).

\bibitem[{\citenamefont{Xie et~al.}(2012)\citenamefont{Xie, Doblhoff-Dier,
  Roither, Sch\"offler, Kartashov, Xu, Rathje, Paulus,
  Baltu\ifmmode~\check{s}\else \v{s}\fi{}ka, Gr\"afe et~al.}}]{Xie12}
\bibinfo{author}{\bibfnamefont{X.}~\bibnamefont{Xie}},
  \bibinfo{author}{\bibfnamefont{K.}~\bibnamefont{Doblhoff-Dier}},
  \bibinfo{author}{\bibfnamefont{S.}~\bibnamefont{Roither}},
  \bibinfo{author}{\bibfnamefont{M.~S.} \bibnamefont{Sch\"offler}},
  \bibinfo{author}{\bibfnamefont{D.}~\bibnamefont{Kartashov}},
  \bibinfo{author}{\bibfnamefont{H.}~\bibnamefont{Xu}},
  \bibinfo{author}{\bibfnamefont{T.}~\bibnamefont{Rathje}},
  \bibinfo{author}{\bibfnamefont{G.~G.} \bibnamefont{Paulus}},
  \bibinfo{author}{\bibfnamefont{A.}~\bibnamefont{Baltu\ifmmode~\check{s}\else
  \v{s}\fi{}ka}}, \bibinfo{author}{\bibfnamefont{S.}~\bibnamefont{Gr\"afe}},
  \bibnamefont{et~al.}, \bibinfo{journal}{Phys. Rev. Lett.}
  \textbf{\bibinfo{volume}{109}}, \bibinfo{pages}{243001}
  (\bibinfo{year}{2012}).

\bibitem[{\citenamefont{Su{\'a}rez et~al.}(2015)\citenamefont{Su{\'a}rez,
  Chac{\'o}n, Ciappina, Biegert, and Lewenstein}}]{Sua15}
\bibinfo{author}{\bibfnamefont{N.}~\bibnamefont{Su{\'a}rez}},
  \bibinfo{author}{\bibfnamefont{A.}~\bibnamefont{Chac{\'o}n}},
  \bibinfo{author}{\bibfnamefont{M.~F.} \bibnamefont{Ciappina}},
  \bibinfo{author}{\bibfnamefont{J.}~\bibnamefont{Biegert}}, \bibnamefont{and}
  \bibinfo{author}{\bibfnamefont{M.}~\bibnamefont{Lewenstein}},
  \bibinfo{journal}{Phys. Rev. A} \textbf{\bibinfo{volume}{92}},
  \bibinfo{pages}{063421} (\bibinfo{year}{2015}).

\bibitem[{\citenamefont{Brabec and Krausz}(2000)}]{Bra00}
\bibinfo{author}{\bibfnamefont{T.}~\bibnamefont{Brabec}} \bibnamefont{and}
  \bibinfo{author}{\bibfnamefont{F.}~\bibnamefont{Krausz}},
  \bibinfo{journal}{Rev. Mod. Phys.} \textbf{\bibinfo{volume}{72}},
  \bibinfo{pages}{545} (\bibinfo{year}{2000}).

\bibitem[{\citenamefont{Becker et~al.}(2002)\citenamefont{Becker, Grasbon,
  Kopold, Milosevic, Paulus, and Walther}}]{Bec02}
\bibinfo{author}{\bibfnamefont{W.}~\bibnamefont{Becker}},
  \bibinfo{author}{\bibfnamefont{F.}~\bibnamefont{Grasbon}},
  \bibinfo{author}{\bibfnamefont{R.}~\bibnamefont{Kopold}},
  \bibinfo{author}{\bibfnamefont{D.}~\bibnamefont{Milosevic}},
  \bibinfo{author}{\bibfnamefont{G.}~\bibnamefont{Paulus}}, \bibnamefont{and}
  \bibinfo{author}{\bibfnamefont{H.}~\bibnamefont{Walther}},
  \bibinfo{journal}{Adv. Atom Mol. Opt. Phys.} \textbf{\bibinfo{volume}{48}},
  \bibinfo{pages}{35} (\bibinfo{year}{2002}).

\bibitem[{\citenamefont{Milo{\v{s}}evi{\'c}
  et~al.}(2006)\citenamefont{Milo{\v{s}}evi{\'c}, Paulus, Bauer, and
  Becker}}]{Mil06}
\bibinfo{author}{\bibfnamefont{D.}~\bibnamefont{Milo{\v{s}}evi{\'c}}},
  \bibinfo{author}{\bibfnamefont{G.}~\bibnamefont{Paulus}},
  \bibinfo{author}{\bibfnamefont{D.}~\bibnamefont{Bauer}}, \bibnamefont{and}
  \bibinfo{author}{\bibfnamefont{W.}~\bibnamefont{Becker}},
  \bibinfo{journal}{J. Phys. B: At. Mol. Opt. Phys.}
  \textbf{\bibinfo{volume}{39}}, \bibinfo{pages}{R203} (\bibinfo{year}{2006}).

\bibitem[{\citenamefont{Haessler et~al.}(2011)\citenamefont{Haessler, Caillat,
  and Salieres}}]{Hae11}
\bibinfo{author}{\bibfnamefont{S.}~\bibnamefont{Haessler}},
  \bibinfo{author}{\bibfnamefont{J.}~\bibnamefont{Caillat}}, \bibnamefont{and}
  \bibinfo{author}{\bibfnamefont{P.}~\bibnamefont{Salieres}},
  \bibinfo{journal}{J. Phys. B: At. Mol. Opt. Phys.}
  \textbf{\bibinfo{volume}{44}}, \bibinfo{pages}{203001}
  (\bibinfo{year}{2011}).

\bibitem[{\citenamefont{Morishita et~al.}(2008)\citenamefont{Morishita, Le,
  Chen, and Lin}}]{Mor08}
\bibinfo{author}{\bibfnamefont{T.}~\bibnamefont{Morishita}},
  \bibinfo{author}{\bibfnamefont{A.-T.} \bibnamefont{Le}},
  \bibinfo{author}{\bibfnamefont{Z.}~\bibnamefont{Chen}}, \bibnamefont{and}
  \bibinfo{author}{\bibfnamefont{C.~D.} \bibnamefont{Lin}},
  \bibinfo{journal}{Phys. Rev. Lett.} \textbf{\bibinfo{volume}{100}},
  \bibinfo{pages}{013903} (\bibinfo{year}{2008}).

\bibitem[{\citenamefont{Kang et~al.}(2010)\citenamefont{Kang, Quan, Wang, Lin,
  Wu, Liu, Liu, Wang, Liu, Gu et~al.}}]{Kan10}
\bibinfo{author}{\bibfnamefont{H.}~\bibnamefont{Kang}},
  \bibinfo{author}{\bibfnamefont{W.}~\bibnamefont{Quan}},
  \bibinfo{author}{\bibfnamefont{Y.}~\bibnamefont{Wang}},
  \bibinfo{author}{\bibfnamefont{Z.}~\bibnamefont{Lin}},
  \bibinfo{author}{\bibfnamefont{M.}~\bibnamefont{Wu}},
  \bibinfo{author}{\bibfnamefont{H.}~\bibnamefont{Liu}},
  \bibinfo{author}{\bibfnamefont{X.}~\bibnamefont{Liu}},
  \bibinfo{author}{\bibfnamefont{B.~B.} \bibnamefont{Wang}},
  \bibinfo{author}{\bibfnamefont{H.~J.} \bibnamefont{Liu}},
  \bibinfo{author}{\bibfnamefont{Y.~Q.} \bibnamefont{Gu}},
  \bibnamefont{et~al.}, \bibinfo{journal}{Phys. Rev. Lett.}
  \textbf{\bibinfo{volume}{104}}, \bibinfo{pages}{203001}
  (\bibinfo{year}{2010}).

\bibitem[{\citenamefont{Paulus et~al.}(2003)\citenamefont{Paulus, Lindner,
  Walther, Baltu\ifmmode~\check{s}\else \v{s}\fi{}ka, Goulielmakis, Lezius, and
  Krausz}}]{Pau03}
\bibinfo{author}{\bibfnamefont{G.~G.} \bibnamefont{Paulus}},
  \bibinfo{author}{\bibfnamefont{F.}~\bibnamefont{Lindner}},
  \bibinfo{author}{\bibfnamefont{H.}~\bibnamefont{Walther}},
  \bibinfo{author}{\bibfnamefont{A.}~\bibnamefont{Baltu\ifmmode~\check{s}\else
  \v{s}\fi{}ka}},
  \bibinfo{author}{\bibfnamefont{E.}~\bibnamefont{Goulielmakis}},
  \bibinfo{author}{\bibfnamefont{M.}~\bibnamefont{Lezius}}, \bibnamefont{and}
  \bibinfo{author}{\bibfnamefont{F.}~\bibnamefont{Krausz}},
  \bibinfo{journal}{Phys. Rev. Lett.} \textbf{\bibinfo{volume}{91}},
  \bibinfo{pages}{253004} (\bibinfo{year}{2003}).

\bibitem[{\citenamefont{Corkum}(1993)}]{Cor93}
\bibinfo{author}{\bibfnamefont{P.~B.} \bibnamefont{Corkum}},
  \bibinfo{journal}{Phys. Rev. Lett.} \textbf{\bibinfo{volume}{71}},
  \bibinfo{pages}{1994} (\bibinfo{year}{1993}).

\bibitem[{\citenamefont{Paulus et~al.}(1994)\citenamefont{Paulus, Becker,
  Nicklich, and Walther}}]{Pau94}
\bibinfo{author}{\bibfnamefont{G.~G.} \bibnamefont{Paulus}},
  \bibinfo{author}{\bibfnamefont{W.}~\bibnamefont{Becker}},
  \bibinfo{author}{\bibfnamefont{W.}~\bibnamefont{Nicklich}}, \bibnamefont{and}
  \bibinfo{author}{\bibfnamefont{H.}~\bibnamefont{Walther}},
  \bibinfo{journal}{J. Phys. B: At. Mol. Opt. Phys.}
  \textbf{\bibinfo{volume}{27}}, \bibinfo{pages}{L703} (\bibinfo{year}{1994}).

\bibitem[{\citenamefont{Milo{\v{s}}evi{\'c}
  et~al.}(2003)\citenamefont{Milo{\v{s}}evi{\'c}, Paulus, and Becker}}]{Mil03}
\bibinfo{author}{\bibfnamefont{D.}~\bibnamefont{Milo{\v{s}}evi{\'c}}},
  \bibinfo{author}{\bibfnamefont{G.}~\bibnamefont{Paulus}}, \bibnamefont{and}
  \bibinfo{author}{\bibfnamefont{W.}~\bibnamefont{Becker}},
  \bibinfo{journal}{Opt. Express} \textbf{\bibinfo{volume}{11}},
  \bibinfo{pages}{1418} (\bibinfo{year}{2003}).

\bibitem[{\citenamefont{Tong et~al.}(2006)\citenamefont{Tong, Hino, and
  Toshima}}]{Ton06}
\bibinfo{author}{\bibfnamefont{X.~M.} \bibnamefont{Tong}},
  \bibinfo{author}{\bibfnamefont{K.}~\bibnamefont{Hino}}, \bibnamefont{and}
  \bibinfo{author}{\bibfnamefont{N.}~\bibnamefont{Toshima}},
  \bibinfo{journal}{Phys. Rev. A} \textbf{\bibinfo{volume}{74}},
  \bibinfo{pages}{031405} (\bibinfo{year}{2006}).

\bibitem[{\citenamefont{Liao et~al.}(2008)\citenamefont{Liao, Lu, Lan, Cao, and
  Li}}]{Qin08}
\bibinfo{author}{\bibfnamefont{Q.}~\bibnamefont{Liao}},
  \bibinfo{author}{\bibfnamefont{P.}~\bibnamefont{Lu}},
  \bibinfo{author}{\bibfnamefont{P.}~\bibnamefont{Lan}},
  \bibinfo{author}{\bibfnamefont{W.}~\bibnamefont{Cao}}, \bibnamefont{and}
  \bibinfo{author}{\bibfnamefont{Y.}~\bibnamefont{Li}}, \bibinfo{journal}{Phys.
  Rev. A} \textbf{\bibinfo{volume}{77}}, \bibinfo{pages}{013408}
  (\bibinfo{year}{2008}).

\bibitem[{\citenamefont{Kling et~al.}(2008)\citenamefont{Kling, Rauschenberger,
  Verhoef, Hasovi{\'c}, Uphues, Milo{\v{s}}evi{\'c}, Muller, and
  Vrakking}}]{Kli08b}
\bibinfo{author}{\bibfnamefont{M.}~\bibnamefont{Kling}},
  \bibinfo{author}{\bibfnamefont{J.}~\bibnamefont{Rauschenberger}},
  \bibinfo{author}{\bibfnamefont{A.}~\bibnamefont{Verhoef}},
  \bibinfo{author}{\bibfnamefont{E.}~\bibnamefont{Hasovi{\'c}}},
  \bibinfo{author}{\bibfnamefont{T.}~\bibnamefont{Uphues}},
  \bibinfo{author}{\bibfnamefont{D.}~\bibnamefont{Milo{\v{s}}evi{\'c}}},
  \bibinfo{author}{\bibfnamefont{H.}~\bibnamefont{Muller}}, \bibnamefont{and}
  \bibinfo{author}{\bibfnamefont{M.}~\bibnamefont{Vrakking}},
  \bibinfo{journal}{New J. Phys.} \textbf{\bibinfo{volume}{10}},
  \bibinfo{pages}{025024} (\bibinfo{year}{2008}).

\bibitem[{\citenamefont{Lindner et~al.}(2005)\citenamefont{Lindner, Sch\"atzel,
  Walther, Baltu\ifmmode~\check{s}\else \v{s}\fi{}ka, Goulielmakis, Krausz,
  Milo\ifmmode \check{s}\else \v{s}\fi{}evi\ifmmode~\acute{c}\else \'{c}\fi{},
  Bauer, Becker, and Paulus}}]{Lin05}
\bibinfo{author}{\bibfnamefont{F.}~\bibnamefont{Lindner}},
  \bibinfo{author}{\bibfnamefont{M.~G.} \bibnamefont{Sch\"atzel}},
  \bibinfo{author}{\bibfnamefont{H.}~\bibnamefont{Walther}},
  \bibinfo{author}{\bibfnamefont{A.}~\bibnamefont{Baltu\ifmmode~\check{s}\else
  \v{s}\fi{}ka}},
  \bibinfo{author}{\bibfnamefont{E.}~\bibnamefont{Goulielmakis}},
  \bibinfo{author}{\bibfnamefont{F.}~\bibnamefont{Krausz}},
  \bibinfo{author}{\bibfnamefont{D.~B.} \bibnamefont{Milo\ifmmode
  \check{s}\else \v{s}\fi{}evi\ifmmode~\acute{c}\else \'{c}\fi{}}},
  \bibinfo{author}{\bibfnamefont{D.}~\bibnamefont{Bauer}},
  \bibinfo{author}{\bibfnamefont{W.}~\bibnamefont{Becker}}, \bibnamefont{and}
  \bibinfo{author}{\bibfnamefont{G.~G.} \bibnamefont{Paulus}},
  \bibinfo{journal}{Phys. Rev. Lett.} \textbf{\bibinfo{volume}{95}},
  \bibinfo{pages}{040401} (\bibinfo{year}{2005}).

\bibitem[{\citenamefont{Gazibegovi\ifmmode \acute{c}\else
  \'{c}\fi{}-Busulad\ifmmode \check{z}\else \v{z}\fi{}i\ifmmode~\acute{c}\else
  \'{c}\fi{} et~al.}(2011)\citenamefont{Gazibegovi\ifmmode \acute{c}\else
  \'{c}\fi{}-Busulad\ifmmode \check{z}\else \v{z}\fi{}i\ifmmode~\acute{c}\else
  \'{c}\fi{}, Hasovi\ifmmode~\acute{c}\else \'{c}\fi{}, Busulad\ifmmode
  \check{z}\else \v{z}\fi{}i\ifmmode~\acute{c}\else \'{c}\fi{}, Milo\ifmmode
  \check{s}\else \v{s}\fi{}evi\ifmmode~\acute{c}\else \'{c}\fi{}, Kelkensberg,
  Siu, Vrakking, L\'epine, Sansone, Nisoli et~al.}}]{Gaz11}
\bibinfo{author}{\bibfnamefont{A.}~\bibnamefont{Gazibegovi\ifmmode
  \acute{c}\else \'{c}\fi{}-Busulad\ifmmode \check{z}\else
  \v{z}\fi{}i\ifmmode~\acute{c}\else \'{c}\fi{}}},
  \bibinfo{author}{\bibfnamefont{E.}~\bibnamefont{Hasovi\ifmmode~\acute{c}\else
  \'{c}\fi{}}}, \bibinfo{author}{\bibfnamefont{M.}~\bibnamefont{Busulad\ifmmode
  \check{z}\else \v{z}\fi{}i\ifmmode~\acute{c}\else \'{c}\fi{}}},
  \bibinfo{author}{\bibfnamefont{D.~B.} \bibnamefont{Milo\ifmmode
  \check{s}\else \v{s}\fi{}evi\ifmmode~\acute{c}\else \'{c}\fi{}}},
  \bibinfo{author}{\bibfnamefont{F.}~\bibnamefont{Kelkensberg}},
  \bibinfo{author}{\bibfnamefont{W.~K.} \bibnamefont{Siu}},
  \bibinfo{author}{\bibfnamefont{M.~J.~J.} \bibnamefont{Vrakking}},
  \bibinfo{author}{\bibfnamefont{F.}~\bibnamefont{L\'epine}},
  \bibinfo{author}{\bibfnamefont{G.}~\bibnamefont{Sansone}},
  \bibinfo{author}{\bibfnamefont{M.}~\bibnamefont{Nisoli}},
  \bibnamefont{et~al.}, \bibinfo{journal}{Phys. Rev. A}
  \textbf{\bibinfo{volume}{84}}, \bibinfo{pages}{043426}
  (\bibinfo{year}{2011}).

\bibitem[{\citenamefont{Kr{\"u}ger et~al.}(2011)\citenamefont{Kr{\"u}ger,
  Schenk, and Hommelhoff}}]{Kru11}
\bibinfo{author}{\bibfnamefont{M.}~\bibnamefont{Kr{\"u}ger}},
  \bibinfo{author}{\bibfnamefont{M.}~\bibnamefont{Schenk}}, \bibnamefont{and}
  \bibinfo{author}{\bibfnamefont{P.}~\bibnamefont{Hommelhoff}},
  \bibinfo{journal}{Nature} \textbf{\bibinfo{volume}{475}}, \bibinfo{pages}{78}
  (\bibinfo{year}{2011}).

\bibitem[{\citenamefont{Park et~al.}(2012)\citenamefont{Park, Piglosiewicz,
  Schmidt, Kollmann, Mascheck, and Lienau}}]{Par12}
\bibinfo{author}{\bibfnamefont{D.~J.} \bibnamefont{Park}},
  \bibinfo{author}{\bibfnamefont{B.}~\bibnamefont{Piglosiewicz}},
  \bibinfo{author}{\bibfnamefont{S.}~\bibnamefont{Schmidt}},
  \bibinfo{author}{\bibfnamefont{H.}~\bibnamefont{Kollmann}},
  \bibinfo{author}{\bibfnamefont{M.}~\bibnamefont{Mascheck}}, \bibnamefont{and}
  \bibinfo{author}{\bibfnamefont{C.}~\bibnamefont{Lienau}},
  \bibinfo{journal}{Phys. Rev. Lett.} \textbf{\bibinfo{volume}{109}},
  \bibinfo{pages}{244803} (\bibinfo{year}{2012}).

\bibitem[{\citenamefont{Kr{\"u}ger et~al.}(2012)\citenamefont{Kr{\"u}ger,
  Schenk, F{\"o}rster, and Hommelhoff}}]{Kru12a}
\bibinfo{author}{\bibfnamefont{M.}~\bibnamefont{Kr{\"u}ger}},
  \bibinfo{author}{\bibfnamefont{M.}~\bibnamefont{Schenk}},
  \bibinfo{author}{\bibfnamefont{M.}~\bibnamefont{F{\"o}rster}},
  \bibnamefont{and}
  \bibinfo{author}{\bibfnamefont{P.}~\bibnamefont{Hommelhoff}},
  \bibinfo{journal}{J. Phys. B: At. Mol. Opt. Phys.}
  \textbf{\bibinfo{volume}{45}}, \bibinfo{pages}{074006}
  (\bibinfo{year}{2012}).

\bibitem[{\citenamefont{Hertel et~al.}(2005)\citenamefont{Hertel, Laarmann, and
  Schulz}}]{Her05}
\bibinfo{author}{\bibfnamefont{I.}~\bibnamefont{Hertel}},
  \bibinfo{author}{\bibfnamefont{T.}~\bibnamefont{Laarmann}}, \bibnamefont{and}
  \bibinfo{author}{\bibfnamefont{C.}~\bibnamefont{Schulz}},
  \bibinfo{journal}{Adv. Atom Mol. Opt. Phys.} \textbf{\bibinfo{volume}{50}},
  \bibinfo{pages}{219} (\bibinfo{year}{2005}).

\bibitem[{\citenamefont{Campbell et~al.}(2006)\citenamefont{Campbell, Hansen,
  Hed{\'e}n, Kjellberg, and Bulgakov}}]{Cam06}
\bibinfo{author}{\bibfnamefont{E.~E.} \bibnamefont{Campbell}},
  \bibinfo{author}{\bibfnamefont{K.}~\bibnamefont{Hansen}},
  \bibinfo{author}{\bibfnamefont{M.}~\bibnamefont{Hed{\'e}n}},
  \bibinfo{author}{\bibfnamefont{M.}~\bibnamefont{Kjellberg}},
  \bibnamefont{and} \bibinfo{author}{\bibfnamefont{A.~V.}
  \bibnamefont{Bulgakov}}, \bibinfo{journal}{Photochem. Photobiol. Sci.}
  \textbf{\bibinfo{volume}{5}}, \bibinfo{pages}{1183} (\bibinfo{year}{2006}).

\bibitem[{\citenamefont{Ganeev}(2011)}]{Gan11}
\bibinfo{author}{\bibfnamefont{R.}~\bibnamefont{Ganeev}},
  \bibinfo{journal}{Laser Phys.} \textbf{\bibinfo{volume}{21}},
  \bibinfo{pages}{25} (\bibinfo{year}{2011}).

\bibitem[{\citenamefont{L{\'e}pine}(2015)}]{Lep15}
\bibinfo{author}{\bibfnamefont{F.}~\bibnamefont{L{\'e}pine}},
  \bibinfo{journal}{J. Phys. B: At. Mol. Opt. Phys.}
  \textbf{\bibinfo{volume}{48}}, \bibinfo{pages}{122002}
  (\bibinfo{year}{2015}).

\bibitem[{\citenamefont{Gao et~al.}(2016)\citenamefont{Gao, Dinh, Kl\"upfel,
  Meier, Reinhard, and Suraud}}]{Gao16}
\bibinfo{author}{\bibfnamefont{C.-Z.} \bibnamefont{Gao}},
  \bibinfo{author}{\bibfnamefont{P.~M.} \bibnamefont{Dinh}},
  \bibinfo{author}{\bibfnamefont{P.}~\bibnamefont{Kl\"upfel}},
  \bibinfo{author}{\bibfnamefont{C.}~\bibnamefont{Meier}},
  \bibinfo{author}{\bibfnamefont{P.-G.} \bibnamefont{Reinhard}},
  \bibnamefont{and} \bibinfo{author}{\bibfnamefont{E.}~\bibnamefont{Suraud}},
  \bibinfo{journal}{Phys. Rev. A} \textbf{\bibinfo{volume}{93}},
  \bibinfo{pages}{022506} (\bibinfo{year}{2016}).

\bibitem[{\citenamefont{Li et~al.}(2015)\citenamefont{Li, Mignolet, Wachter,
  Skruszewicz, Zherebtsov, S{\"u}{\ss}mann, Kessel, Trushin, Kling, K{\"u}bel
  et~al.}}]{Li15}
\bibinfo{author}{\bibfnamefont{H.}~\bibnamefont{Li}},
  \bibinfo{author}{\bibfnamefont{B.}~\bibnamefont{Mignolet}},
  \bibinfo{author}{\bibfnamefont{G.}~\bibnamefont{Wachter}},
  \bibinfo{author}{\bibfnamefont{S.}~\bibnamefont{Skruszewicz}},
  \bibinfo{author}{\bibfnamefont{S.}~\bibnamefont{Zherebtsov}},
  \bibinfo{author}{\bibfnamefont{F.}~\bibnamefont{S{\"u}{\ss}mann}},
  \bibinfo{author}{\bibfnamefont{A.}~\bibnamefont{Kessel}},
  \bibinfo{author}{\bibfnamefont{S.}~\bibnamefont{Trushin}},
  \bibinfo{author}{\bibfnamefont{N.~G.} \bibnamefont{Kling}},
  \bibinfo{author}{\bibfnamefont{M.}~\bibnamefont{K{\"u}bel}},
  \bibnamefont{et~al.}, \bibinfo{journal}{Phys. Rev. Lett.}
  \textbf{\bibinfo{volume}{114}}, \bibinfo{pages}{123004}
  (\bibinfo{year}{2015}).

\bibitem[{\citenamefont{Runge and Gross}(1984)}]{Run84}
\bibinfo{author}{\bibfnamefont{E.}~\bibnamefont{Runge}} \bibnamefont{and}
  \bibinfo{author}{\bibfnamefont{E.~K.} \bibnamefont{Gross}},
  \bibinfo{journal}{Phys. Rev. Lett.} \textbf{\bibinfo{volume}{52}},
  \bibinfo{pages}{997} (\bibinfo{year}{1984}).

\bibitem[{\citenamefont{Dreizler and Gross}(1990)}]{Dre90}
\bibinfo{author}{\bibfnamefont{R.~M.} \bibnamefont{Dreizler}} \bibnamefont{and}
  \bibinfo{author}{\bibfnamefont{E.~K.~U.} \bibnamefont{Gross}},
  \emph{\bibinfo{title}{{Density Functional Theory: An Approach to the Quantum
  Many-Body Problem}}} (\bibinfo{publisher}{Springer-Verlag},
  \bibinfo{address}{Berlin}, \bibinfo{year}{1990}).

\bibitem[{\citenamefont{Perdew and Wang}(1992)}]{Per92}
\bibinfo{author}{\bibfnamefont{J.~P.} \bibnamefont{Perdew}} \bibnamefont{and}
  \bibinfo{author}{\bibfnamefont{Y.}~\bibnamefont{Wang}},
  \bibinfo{journal}{Phys. Rev. B} \textbf{\bibinfo{volume}{45}},
  \bibinfo{pages}{13244} (\bibinfo{year}{1992}).

\bibitem[{\citenamefont{Perdew and Zunger}(1981)}]{PeZ81}
\bibinfo{author}{\bibfnamefont{J.~P.} \bibnamefont{Perdew}} \bibnamefont{and}
  \bibinfo{author}{\bibfnamefont{A.}~\bibnamefont{Zunger}},
  \bibinfo{journal}{Phys. Rev. B} \textbf{\bibinfo{volume}{23}},
  \bibinfo{pages}{5048} (\bibinfo{year}{1981}).

\bibitem[{\citenamefont{Messud et~al.}(2008)\citenamefont{Messud, Dinh,
  Reinhard, and Suraud}}]{Mes08b}
\bibinfo{author}{\bibfnamefont{J.}~\bibnamefont{Messud}},
  \bibinfo{author}{\bibfnamefont{P.~M.} \bibnamefont{Dinh}},
  \bibinfo{author}{\bibfnamefont{P.-G.} \bibnamefont{Reinhard}},
  \bibnamefont{and} \bibinfo{author}{\bibfnamefont{E.}~\bibnamefont{Suraud}},
  \bibinfo{journal}{Ann. Phys. (N.Y.)} \textbf{\bibinfo{volume}{324}},
  \bibinfo{pages}{955} (\bibinfo{year}{2008}).

\bibitem[{\citenamefont{Legrand et~al.}(2002)\citenamefont{Legrand, Suraud, and
  Reinhard}}]{Leg02}
\bibinfo{author}{\bibfnamefont{C.}~\bibnamefont{Legrand}},
  \bibinfo{author}{\bibfnamefont{E.}~\bibnamefont{Suraud}}, \bibnamefont{and}
  \bibinfo{author}{\bibfnamefont{P.-G.} \bibnamefont{Reinhard}},
  \bibinfo{journal}{J. Phys. B: At. Mol. Opt. Phys.}
  \textbf{\bibinfo{volume}{35}}, \bibinfo{pages}{1115} (\bibinfo{year}{2002}).

\bibitem[{\citenamefont{Kl\"upfel et~al.}(2013)\citenamefont{Kl\"upfel, Dinh,
  Reinhard, and Suraud}}]{Klu13a}
\bibinfo{author}{\bibfnamefont{P.}~\bibnamefont{Kl\"upfel}},
  \bibinfo{author}{\bibfnamefont{P.~M.} \bibnamefont{Dinh}},
  \bibinfo{author}{\bibfnamefont{P.-G.} \bibnamefont{Reinhard}},
  \bibnamefont{and} \bibinfo{author}{\bibfnamefont{E.}~\bibnamefont{Suraud}},
  \bibinfo{journal}{Phys. Rev. A} \textbf{\bibinfo{volume}{88}},
  \bibinfo{pages}{052501} (\bibinfo{year}{2013}).

\bibitem[{\citenamefont{M{\"u}ller and Fedorov}(1996)}]{Mul96}
\bibinfo{author}{\bibfnamefont{H.-G.} \bibnamefont{M{\"u}ller}}
  \bibnamefont{and} \bibinfo{author}{\bibfnamefont{M.}~\bibnamefont{Fedorov}},
  \emph{\bibinfo{title}{Super-intense laser-atom physics IV}},
  vol.~\bibinfo{volume}{13} (\bibinfo{publisher}{Springer Science \& Business
  Media}, \bibinfo{year}{1996}).

\bibitem[{\citenamefont{Puska and Nieminen}(1993)}]{Pus93}
\bibinfo{author}{\bibfnamefont{M.}~\bibnamefont{Puska}} \bibnamefont{and}
  \bibinfo{author}{\bibfnamefont{R.~M.} \bibnamefont{Nieminen}},
  \bibinfo{journal}{Phys. Rev. A} \textbf{\bibinfo{volume}{47}},
  \bibinfo{pages}{1181} (\bibinfo{year}{1993}).

\bibitem[{\citenamefont{Bauer et~al.}(2001)\citenamefont{Bauer, Ceccherini,
  Macchi, and Cornolti}}]{Bau01}
\bibinfo{author}{\bibfnamefont{D.}~\bibnamefont{Bauer}},
  \bibinfo{author}{\bibfnamefont{F.}~\bibnamefont{Ceccherini}},
  \bibinfo{author}{\bibfnamefont{A.}~\bibnamefont{Macchi}}, \bibnamefont{and}
  \bibinfo{author}{\bibfnamefont{F.}~\bibnamefont{Cornolti}},
  \bibinfo{journal}{Phys. Rev. A} \textbf{\bibinfo{volume}{64}}
  (\bibinfo{year}{2001}).

\bibitem[{\citenamefont{Cormier et~al.}(2003)\citenamefont{Cormier, Hervieux,
  Wiehle, Witzel, and Helm}}]{Cor03}
\bibinfo{author}{\bibfnamefont{E.}~\bibnamefont{Cormier}},
  \bibinfo{author}{\bibfnamefont{P.-A.} \bibnamefont{Hervieux}},
  \bibinfo{author}{\bibfnamefont{R.}~\bibnamefont{Wiehle}},
  \bibinfo{author}{\bibfnamefont{B.}~\bibnamefont{Witzel}}, \bibnamefont{and}
  \bibinfo{author}{\bibfnamefont{H.}~\bibnamefont{Helm}},
  \bibinfo{journal}{Eur. Phys. J. D} \textbf{\bibinfo{volume}{26}},
  \bibinfo{pages}{83} (\bibinfo{year}{2003}).

\bibitem[{\citenamefont{Reinhard et~al.}(2013)\citenamefont{Reinhard, Wopperer,
  Dinh, and Suraud}}]{Rei13}
\bibinfo{author}{\bibfnamefont{P.-G.} \bibnamefont{Reinhard}},
  \bibinfo{author}{\bibfnamefont{P.}~\bibnamefont{Wopperer}},
  \bibinfo{author}{\bibfnamefont{P.~M.} \bibnamefont{Dinh}}, \bibnamefont{and}
  \bibinfo{author}{\bibfnamefont{E.}~\bibnamefont{Suraud}}, in
  \emph{\bibinfo{booktitle}{ICQNM 2013, The Seventh International Conference on
  Quantum, Nano and Micro Technologies}} (\bibinfo{year}{2013}), pp.
  \bibinfo{pages}{13--17}.

\bibitem[{\citenamefont{Hedberg et~al.}(1991)\citenamefont{Hedberg, Hedberg,
  Bethune, Brown, Dorn, Johnson, and de~Vries}}]{Hed91}
\bibinfo{author}{\bibfnamefont{K.}~\bibnamefont{Hedberg}},
  \bibinfo{author}{\bibfnamefont{L.}~\bibnamefont{Hedberg}},
  \bibinfo{author}{\bibfnamefont{D.~S.} \bibnamefont{Bethune}},
  \bibinfo{author}{\bibfnamefont{C.~A.} \bibnamefont{Brown}},
  \bibinfo{author}{\bibfnamefont{H.~C.} \bibnamefont{Dorn}},
  \bibinfo{author}{\bibfnamefont{R.~D.} \bibnamefont{Johnson}},
  \bibnamefont{and} \bibinfo{author}{\bibfnamefont{M.}~\bibnamefont{de~Vries}},
  \bibinfo{journal}{Science} \textbf{\bibinfo{volume}{254}},
  \bibinfo{pages}{410} (\bibinfo{year}{1991}).

\bibitem[{\citenamefont{Lichtenberger et~al.}(1990)\citenamefont{Lichtenberger,
  Jatcko, Nebesny, Ray, Huffman, and Lamb}}]{Lic91b}
\bibinfo{author}{\bibfnamefont{D.~L.} \bibnamefont{Lichtenberger}},
  \bibinfo{author}{\bibfnamefont{M.~E.} \bibnamefont{Jatcko}},
  \bibinfo{author}{\bibfnamefont{K.~W.} \bibnamefont{Nebesny}},
  \bibinfo{author}{\bibfnamefont{C.~D.} \bibnamefont{Ray}},
  \bibinfo{author}{\bibfnamefont{D.~R.} \bibnamefont{Huffman}},
  \bibnamefont{and} \bibinfo{author}{\bibfnamefont{L.~D.} \bibnamefont{Lamb}},
  \bibinfo{journal}{Mater. Res. Soc. Symp. Proc.}
  \textbf{\bibinfo{volume}{206}}, \bibinfo{pages}{673} (\bibinfo{year}{1990}).

\bibitem[{\citenamefont{Sattler}(2010)}]{Sat10}
\bibinfo{author}{\bibfnamefont{K.}~\bibnamefont{Sattler}},
  \emph{\bibinfo{title}{Handbook of Nanophysics: Clusters and Fullerenes}},
  Handbook of Nanophysics (\bibinfo{publisher}{CRC Press},
  \bibinfo{year}{2010}).

\bibitem[{\citenamefont{Ashcroft and Mermin}(1976)}]{Ash76}
\bibinfo{author}{\bibfnamefont{N.~W.} \bibnamefont{Ashcroft}} \bibnamefont{and}
  \bibinfo{author}{\bibfnamefont{N.~D.} \bibnamefont{Mermin}},
  \emph{\bibinfo{title}{{Solid State Physics}}} (\bibinfo{publisher}{Saunders
  College}, \bibinfo{address}{Philadelphia}, \bibinfo{year}{1976}).

\bibitem[{\citenamefont{Lemell et~al.}(2003)\citenamefont{Lemell, Tong, Krausz,
  and Burgd\"orfer}}]{Lem03}
\bibinfo{author}{\bibfnamefont{C.}~\bibnamefont{Lemell}},
  \bibinfo{author}{\bibfnamefont{X.-M.} \bibnamefont{Tong}},
  \bibinfo{author}{\bibfnamefont{F.}~\bibnamefont{Krausz}}, \bibnamefont{and}
  \bibinfo{author}{\bibfnamefont{J.}~\bibnamefont{Burgd\"orfer}},
  \bibinfo{journal}{Phys. Rev. Lett.} \textbf{\bibinfo{volume}{90}},
  \bibinfo{pages}{076403} (\bibinfo{year}{2003}).

\bibitem[{\citenamefont{Kreibig and Vollmer}(2013)}]{Kre93}
\bibinfo{author}{\bibfnamefont{U.}~\bibnamefont{Kreibig}} \bibnamefont{and}
  \bibinfo{author}{\bibfnamefont{M.}~\bibnamefont{Vollmer}},
  \emph{\bibinfo{title}{Optical properties of metal clusters}},
  vol.~\bibinfo{volume}{25} (\bibinfo{publisher}{Springer Science \& Business
  Media}, \bibinfo{year}{2013}).

\bibitem[{\citenamefont{Brack}(1993)}]{Bra93}
\bibinfo{author}{\bibfnamefont{M.}~\bibnamefont{Brack}}, \bibinfo{journal}{Rev.
  Mod. Phys.} \textbf{\bibinfo{volume}{65}}, \bibinfo{pages}{677}
  (\bibinfo{year}{1993}).

\bibitem[{\citenamefont{Montag and Reinhard}(1995)}]{Mon95a}
\bibinfo{author}{\bibfnamefont{B.}~\bibnamefont{Montag}} \bibnamefont{and}
  \bibinfo{author}{\bibfnamefont{P.-G.} \bibnamefont{Reinhard}},
  \bibinfo{journal}{Z. Phys. D: At., Mol. Clusters}
  \textbf{\bibinfo{volume}{33}}, \bibinfo{pages}{265} (\bibinfo{year}{1995}).

\bibitem[{\citenamefont{Reinhard and Cusson}(1982)}]{Rei82}
\bibinfo{author}{\bibfnamefont{P.-G.} \bibnamefont{Reinhard}} \bibnamefont{and}
  \bibinfo{author}{\bibfnamefont{R.}~\bibnamefont{Cusson}},
  \bibinfo{journal}{Nucl. Phys. A} \textbf{\bibinfo{volume}{378}},
  \bibinfo{pages}{418} (\bibinfo{year}{1982}).

\bibitem[{\citenamefont{Feit et~al.}(1982)\citenamefont{Feit, Fleck, and
  Steiger}}]{Fei82}
\bibinfo{author}{\bibfnamefont{M.}~\bibnamefont{Feit}},
  \bibinfo{author}{\bibfnamefont{J.}~\bibnamefont{Fleck}}, \bibnamefont{and}
  \bibinfo{author}{\bibfnamefont{A.}~\bibnamefont{Steiger}},
  \bibinfo{journal}{J. Comput. Phys.} \textbf{\bibinfo{volume}{47}},
  \bibinfo{pages}{412} (\bibinfo{year}{1982}).

\bibitem[{\citenamefont{Calvayrac et~al.}(2000)\citenamefont{Calvayrac,
  Reinhard, Suraud, and Ullrich}}]{Cal00}
\bibinfo{author}{\bibfnamefont{F.}~\bibnamefont{Calvayrac}},
  \bibinfo{author}{\bibfnamefont{P.-G.} \bibnamefont{Reinhard}},
  \bibinfo{author}{\bibfnamefont{E.}~\bibnamefont{Suraud}}, \bibnamefont{and}
  \bibinfo{author}{\bibfnamefont{C.~A.} \bibnamefont{Ullrich}},
  \bibinfo{journal}{Phys. Rep.} \textbf{\bibinfo{volume}{337}},
  \bibinfo{pages}{493} (\bibinfo{year}{2000}).

\bibitem[{\citenamefont{Reinhard and Suraud}(2003)}]{Rei03a}
\bibinfo{author}{\bibfnamefont{P.-G.} \bibnamefont{Reinhard}} \bibnamefont{and}
  \bibinfo{author}{\bibfnamefont{E.}~\bibnamefont{Suraud}},
  \emph{\bibinfo{title}{Introduction to Cluster Dynamics}}
  (\bibinfo{publisher}{Wiley}, \bibinfo{address}{New York},
  \bibinfo{year}{2003}).

\bibitem[{\citenamefont{Wopperer et~al.}(2015)\citenamefont{Wopperer, Dinh,
  Reinhard, and Suraud}}]{Wop15}
\bibinfo{author}{\bibfnamefont{P.}~\bibnamefont{Wopperer}},
  \bibinfo{author}{\bibfnamefont{P.~M.} \bibnamefont{Dinh}},
  \bibinfo{author}{\bibfnamefont{P.-G.} \bibnamefont{Reinhard}},
  \bibnamefont{and} \bibinfo{author}{\bibfnamefont{E.}~\bibnamefont{Suraud}},
  \bibinfo{journal}{Phys. Rep.} \textbf{\bibinfo{volume}{562}},
  \bibinfo{pages}{1} (\bibinfo{year}{2015}).

\bibitem[{\citenamefont{Reinhard et~al.}(2006)\citenamefont{Reinhard,
  Stevenson, Almehed, Maruhn, and Strayer}}]{PGR06}
\bibinfo{author}{\bibfnamefont{P.-G.} \bibnamefont{Reinhard}},
  \bibinfo{author}{\bibfnamefont{P.~D.} \bibnamefont{Stevenson}},
  \bibinfo{author}{\bibfnamefont{D.}~\bibnamefont{Almehed}},
  \bibinfo{author}{\bibfnamefont{J.~A.} \bibnamefont{Maruhn}},
  \bibnamefont{and} \bibinfo{author}{\bibfnamefont{M.~R.}
  \bibnamefont{Strayer}}, \bibinfo{journal}{Phys. Rev. E}
  \textbf{\bibinfo{volume}{73}}, \bibinfo{pages}{036709}
  (\bibinfo{year}{2006}).

\bibitem[{\citenamefont{Pohl et~al.}(2001)\citenamefont{Pohl, Reinhard, and
  Suraud}}]{Poh01}
\bibinfo{author}{\bibfnamefont{A.}~\bibnamefont{Pohl}},
  \bibinfo{author}{\bibfnamefont{P.-G.} \bibnamefont{Reinhard}},
  \bibnamefont{and} \bibinfo{author}{\bibfnamefont{E.}~\bibnamefont{Suraud}},
  \bibinfo{journal}{J. Phys. B} \textbf{\bibinfo{volume}{34}},
  \bibinfo{pages}{4969} (\bibinfo{year}{2001}).

\bibitem[{\citenamefont{De~Giovannini et~al.}(2012)\citenamefont{De~Giovannini,
  Varsano, Marques, Appel, Gross, and Rubio}}]{DeG12}
\bibinfo{author}{\bibfnamefont{U.}~\bibnamefont{De~Giovannini}},
  \bibinfo{author}{\bibfnamefont{D.}~\bibnamefont{Varsano}},
  \bibinfo{author}{\bibfnamefont{M.~A.~L.} \bibnamefont{Marques}},
  \bibinfo{author}{\bibfnamefont{H.}~\bibnamefont{Appel}},
  \bibinfo{author}{\bibfnamefont{E.~K.~U.} \bibnamefont{Gross}},
  \bibnamefont{and} \bibinfo{author}{\bibfnamefont{A.}~\bibnamefont{Rubio}},
  \bibinfo{journal}{Phys. Rev. A} \textbf{\bibinfo{volume}{85}},
  \bibinfo{pages}{062515} (\bibinfo{year}{2012}).

\bibitem[{\citenamefont{Dinh et~al.}(2013)\citenamefont{Dinh, Romaniello,
  Reinhard, and Suraud}}]{Din13}
\bibinfo{author}{\bibfnamefont{P.~M.} \bibnamefont{Dinh}},
  \bibinfo{author}{\bibfnamefont{P.}~\bibnamefont{Romaniello}},
  \bibinfo{author}{\bibfnamefont{P.-G.} \bibnamefont{Reinhard}},
  \bibnamefont{and} \bibinfo{author}{\bibfnamefont{E.}~\bibnamefont{Suraud}},
  \bibinfo{journal}{Phys. Rev. A} \textbf{\bibinfo{volume}{87}},
  \bibinfo{pages}{032514} (\bibinfo{year}{2013}).

\bibitem[{\citenamefont{Dauth and K\"ummel}(2016)}]{Dau16}
\bibinfo{author}{\bibfnamefont{M.}~\bibnamefont{Dauth}} \bibnamefont{and}
  \bibinfo{author}{\bibfnamefont{S.}~\bibnamefont{K\"ummel}},
  \bibinfo{journal}{Phys. Rev. A} \textbf{\bibinfo{volume}{93}},
  \bibinfo{pages}{022502} (\bibinfo{year}{2016}).

\bibitem[{\citenamefont{Giulietti et~al.}(2006)\citenamefont{Giulietti,
  Tomassini, Galimberti, Gizzi, Koester, Labate, Ceccotti, D’Oliveira,
  Auguste, Monot et~al.}}]{Giu06}
\bibinfo{author}{\bibfnamefont{A.}~\bibnamefont{Giulietti}},
  \bibinfo{author}{\bibfnamefont{P.}~\bibnamefont{Tomassini}},
  \bibinfo{author}{\bibfnamefont{M.}~\bibnamefont{Galimberti}},
  \bibinfo{author}{\bibfnamefont{D.~G. L.~A.} \bibnamefont{Gizzi}},
  \bibinfo{author}{\bibfnamefont{P.}~\bibnamefont{Koester}},
  \bibinfo{author}{\bibfnamefont{L.}~\bibnamefont{Labate}},
  \bibinfo{author}{\bibfnamefont{T.}~\bibnamefont{Ceccotti}},
  \bibinfo{author}{\bibfnamefont{P.}~\bibnamefont{D’Oliveira}},
  \bibinfo{author}{\bibfnamefont{T.}~\bibnamefont{Auguste}},
  \bibinfo{author}{\bibfnamefont{P.}~\bibnamefont{Monot}},
  \bibnamefont{et~al.}, \bibinfo{journal}{Phys. Plasmas}
  \textbf{\bibinfo{volume}{13}}, \bibinfo{pages}{093103}
  (\bibinfo{year}{2006}).

\bibitem[{\citenamefont{Paulus et~al.}(1995)\citenamefont{Paulus, Becker, and
  Walther}}]{Pau95}
\bibinfo{author}{\bibfnamefont{G.}~\bibnamefont{Paulus}},
  \bibinfo{author}{\bibfnamefont{W.}~\bibnamefont{Becker}}, \bibnamefont{and}
  \bibinfo{author}{\bibfnamefont{H.}~\bibnamefont{Walther}},
  \bibinfo{journal}{Phys. Rev. A} \textbf{\bibinfo{volume}{52}},
  \bibinfo{pages}{4043} (\bibinfo{year}{1995}).

\bibitem[{\citenamefont{Nguyen et~al.}(2004)\citenamefont{Nguyen, Bandrauk, and
  Ullrich}}]{Ngu04}
\bibinfo{author}{\bibfnamefont{H.}~\bibnamefont{Nguyen}},
  \bibinfo{author}{\bibfnamefont{A.}~\bibnamefont{Bandrauk}}, \bibnamefont{and}
  \bibinfo{author}{\bibfnamefont{C.~A.} \bibnamefont{Ullrich}},
  \bibinfo{journal}{Phys. Rev. A} \textbf{\bibinfo{volume}{69}},
  \bibinfo{pages}{063415} (\bibinfo{year}{2004}).

\bibitem[{\citenamefont{Skruszewicz et~al.}(2015)\citenamefont{Skruszewicz,
  Tiggesb\"aumker, Meiwes-Broer, Arbeiter, Fennel, and Bauer}}]{Skr15}
\bibinfo{author}{\bibfnamefont{S.}~\bibnamefont{Skruszewicz}},
  \bibinfo{author}{\bibfnamefont{J.}~\bibnamefont{Tiggesb\"aumker}},
  \bibinfo{author}{\bibfnamefont{K.-H.} \bibnamefont{Meiwes-Broer}},
  \bibinfo{author}{\bibfnamefont{M.}~\bibnamefont{Arbeiter}},
  \bibinfo{author}{\bibfnamefont{T.}~\bibnamefont{Fennel}}, \bibnamefont{and}
  \bibinfo{author}{\bibfnamefont{D.}~\bibnamefont{Bauer}},
  \bibinfo{journal}{Phys. Rev. Lett.} \textbf{\bibinfo{volume}{115}},
  \bibinfo{pages}{043001} (\bibinfo{year}{2015}).

\end{thebibliography}

\end{document}